\documentclass[conference]{IEEEtran}
\pdfoutput=1
\def\ifextended#1#2{#1}

\usepackage[utf8]{inputenc}
\usepackage[pdftex]{color,graphicx}
\usepackage[english]{babel}
\usepackage[noadjust,nocompress]{cite}
\usepackage[caption=false]{subfig}
\usepackage{xspace}
\usepackage{subscript}
\usepackage{tikz}
\usetikzlibrary{patterns}
\usepackage{tabto}
\usepackage{amsthm}
\usepackage{latexsym}
\usepackage{dsfont}
\usepackage{hyperref}

\newcommand{\system}[0]{\textsc{Isos}\xspace}

\newcommand{\headline}[1]{\vspace{.5mm}\noindent\textbf{\textit{#1.}}~}

\usepackage{listings}
\usepackage{subscript}

\font\lsttt=rm-lmtl10 scaled 820
\font\lstbtt=rm-lmtk10 scaled 820

\newcommand{\commentsize}{\fontsize{8pt}{0pt}\selectfont}
\newcommand{\pseudocode}{\commentsize\normalfont}

\hypersetup{
	pdfsubject = {Egalitarian Byzantine Fault Tolerance},
	pdftitle = {Egalitarian Byzantine Fault Tolerance},
	pdfauthor = {Michael Eischer and Tobias Distler},
	pdfkeywords = {State-Machine Replication, Byzantine Fault Tolerance, Geo-Replication, Leaderless Consensus},
	pdfborder = {0 0 0}
}

\title{Egalitarian Byzantine Fault Tolerance\\(Extended Version)\vspace{-0.35cm}}

\author{
	\IEEEauthorblockN{Michael Eischer, and Tobias Distler}
	\IEEEauthorblockA{Friedrich-Alexander University Erlangen-N\"urnberg (FAU)\\Email: \{eischer,distler\}@cs.fau.de\vspace{-0.35cm}}
}

\usepackage[firstpage=true,placement=bottom]{background}
\newcommand{\disclaimer}{
\begin{tikzpicture}[remember picture, overlay, black]
\node{
\begin{minipage}{\textwidth}
\setlength{\fboxsep}{1em}%
\newlength{\copylen}\setlength{\copylen}{\textwidth}\addtolength{\copylen}{-2\fboxsep}%
\colorbox{yellow!70!black!20}{%
\parbox{\copylen}{%
\sffamily\footnotesize\centering%
This is an extended version of the article "Michael Eischer and Tobias Distler. 2021. Egalitarian Byzantine Fault Tolerance. In \emph{Proceedings of the 26th Pacific Rim International Symposium on Dependable Computing (PRDC\,'21), Perth, Australia, 1--4 December 2021.}"
}}
\vspace{4mm}
\end{minipage}
};
\end{tikzpicture}
}
\SetBgContents{\disclaimer}
\SetBgOpacity{1.0}
\SetBgAngle{0.0}
\SetBgScale{1.0}

\begin{document}
\maketitle

\begin{abstract}

Minimizing end-to-end latency in geo-re\-pli\-ca\-ted systems usually makes it necessary to compromise on resilience, resource efficiency, or throughput performance, because existing approaches either tolerate only crashes, require additional replicas, or rely on a global leader for consensus. In this paper, we eliminate the need for such tradeoffs by presenting \system, a leaderless replication protocol that tolerates up to $f$~Byzantine faults with a minimum of $3f+1$~replicas. To reduce latency in wide-area environments, \system relies on an efficient consensus algorithm that allows all participating replicas to propose new requests and thereby enables clients to avoid delays by submitting requests to their nearest replica. In addition, \system minimizes overhead by limiting message ordering to requests that conflict with each other~(e.g.,~due to accessing the same state parts) and by already committing them after three communication steps if at least $f+1$~replicas report each conflict. Our experimental evaluation with a geo-replicated key-value store shows that these properties allow \system to provide lower end-to-end latency than existing protocols, especially for use-case scenarios in which the clients of a system are distributed across multiple locations.

\end{abstract}


\def\msgreq{\textsc{Req}\xspace}
\def\msgpropose{\textsc{Dep\-Propose}\xspace}
\def\msgverify{\textsc{Dep\-Verify}\xspace}
\def\msgfpcommit{\textsc{Dep\-Commit}\xspace}
\def\msgpreprepare{\textsc{Preprepare}\xspace}
\def\msgprepare{\textsc{Prepare}\xspace}
\def\msgcommit{\textsc{Commit}\xspace}
\def\msgvc{\textsc{View\-Change}\xspace}
\def\msgnewview{\textsc{New\-View}\xspace}
\def\reqa{$A$\xspace}
\def\reqb{$B$\xspace}
\def\msgcp{\textsc{Checkpoint}\xspace}
\def\msgqueryexec{\textsc{Query\-Exec}\xspace}
\def\msgexec{\textsc{Exec}\xspace}
\def\msgcpreq{\textsc{CpReq}\xspace}

\section{Introduction}
\label{sec:introduction}

Distributing a replicated service across several geographic sites offers the possibility to make the service resilient against a wide spectrum of faults, including failures of entire data centers. Unfortunately, traditional state-machine replication approaches~\cite{schneider90implementing,castro99practical} in such environments incur high latency due to electing a leader replica which is then responsible for establishing a total order on all incoming client requests. Relying on a single global leader replica in wide-area environments comes with the major drawbacks of (1)~creating a potential performance bottleneck, (2)~disadvantaging clients that reside at a greater distance to the current leader, and (3)~introducing response-time volatility, because overall latency can vary significantly depending on where the acting leader is located. Although it is possible to rotate the leader role among replicas~\cite{mao08mencius}, this technique only slightly mitigates the problem since the rotation process itself introduces coordination overhead in the form of (at least) an additional communication step.

Several existing works~\cite{abd-el-malek05qu,moraru13there,enes20state,pires18generalized,bazzi21clairvoyant} address these issues by building on the insight that for guaranteeing linearizability~\cite{herlihy90linearizability} it is not actually necessary to totally order all client requests that are submitted to a service. Instead, the efficiency of message ordering in many cases can be improved by taking the semantics of requests into account~\cite{pedone99generic} and only ordering those requests that conflict with each other, for example due to operating on the same application-state variables. In recent years, applications of this principle led to a variety of protocols that explore different points in the design space of replicated systems. Specifically, this includes protocols that have been designed to tolerate crashes~\cite{moraru13there,enes20state}, Byzantine fault-tolerant~(BFT) protocols achieving efficiency at the cost of additional replicas~\cite{abd-el-malek05qu,bazzi21clairvoyant}, as well as protocols that rely on a global leader replica to ensure progress in case of disagreements between different replicas~\cite{pires18generalized}. While on the one hand illustrating the effectiveness and flexibility of the underlying concept, this variety of protocols on the other hand also means that existing approaches require compromising on resilience, resource efficiency, or throughput performance.

To eliminate the need for such tradeoffs, our goal was to develop a protocol that combines all three desirable properties while still providing low latency.
The result of our efforts is \system, a state-machine replication protocol that tolerates Byzantine faults, demands only the minimum group size necessary for BFT in asynchronous environments~(i.e.,~\mbox{$3f+1$} replicas to tolerate $f$~faults), and operates without global leader replica. To minimize end-to-end latency in geo-replicated settings, \system offers a fast path that enables replicas to execute client requests after three consensus communication steps if either (a)~there currently are no conflicting requests or (b)~each conflict is identified by at least $f+1$~replicas. In the (typically rare) case in which none of the two scenarios applies, \system switches to a fallback path that is then responsible for resolving the discrepancies between replicas. Since neither of the two paths in \system requires the election of a global leader, we \linebreak refer to this concept as \emph{egalitarian} Byzantine fault tolerance.

In summary, this paper makes the following contributions: (1)~It presents \system's efficient BFT consensus algorithm that only orders conflicting requests and avoids a global leader during both normal-case operation as well as conflict-discrepancy resolution. (2)~It shows how \system's request-execution stage is able to safely operate with a bounded state, and this despite the fact that faulty replicas possibly introduce request-dependency chains of infinite length.
(3)~It details \system's checkpointing mechanism that enables the protocol to garbage-collect consensus information about already ordered requests; garbage collection is a relevant problem in practice, but often not implemented in other protocols~(e.g.,~EPaxos~\cite{moraru13there}). (4)~It formally proves the correctness of both \system's agreement and execution stage. Notice that due to space limitations, we limit Section~\ref{sec:dependency-correctness} to the presentation of a proof sketch; the full proof (as well as a pseudocode summary of \system's agreement protocol) is available in \ifextended{Appendix~\ref{sec:appendix-pseudo-code}}{the extended version of this paper~\cite{isos-extended}}. (5)~It experimentally evaluates \system with a key-value store in a geo-distributed setting deployed in Amazon's EC2 cloud.

\section{System Model}
\label{sec:model}

In this work we focus on stateful applications that are replicated across multiple servers for fault tolerance. To remain available even in the presence of data-center outages, the replicas of a system are hosted at different geographic sites, as illustrated in Figure~\ref{fig:background}. Clients of the service typically reside in proximity to one of the replicas, often within the same data center. As a result of such a setting, overall response times in our target systems are dominated by the latency induced by the state-machine replication protocol executed between servers.

We assume that the replicated service must provide safety in the presence of Byzantine faults as well as an asynchronous network. To further be able to ensure liveness despite the FLP impossibility~\cite{fischer85impossibility}, there need to be synchronous phases during which the one-way network delay between all pairs of replicas is below a threshold~$\Delta$, which is known to replicas.
Clients and replicas communicate over the network by exchanging messages that are signed with the sender's private key, denoted as $\langle ... \rangle _{\sigma_i}$ for a sender~$i$. Recipients immediately discard \linebreak messages in case they are unable to verify the signature.

Clients invoke operations in the application by submitting requests to the server side. With regard to the execution of requests, we define a predicate $con\hspace{-.2mm}flict(a, b)$ which holds if there is an interdependency between two requests $a$ and~$b$. Specifically, two requests are in conflict with each other if their effects~(i.e.,~changes to the application state) and outcomes~(i.e.,~results) vary depending on the relative order in which they are executed by a replica. In addition, we define that $con\hspace{-.2mm}flict(a, b)$ always holds for requests issued by the same client. Several previous works~\cite{kotla04high,distler11increasing,moraru13there,li16sarek,enes20state,bazzi21clairvoyant} relied on similar predicates and concluded that for many applications determining request conflicts is straightforward. In key-value stores, for example, requests typically contain the key(s) of the data set(s) they access. Consequently, a write can be identified to conflict with another write or read to the same key. In contrast, two reads of the same data set are independent of each other due to not modifying application state and their results not being influenced by their relative execution order.

With our work presented in this paper we target use-case scenarios in which conflicting requests only constitute a small fraction of the application's overall workload~(e.g.,~less than 5\%~\cite{moraru13there}). In practice, this for example is the case for key-value stores with high read-to-write ratios~\cite{xu13characterizing} or coordination services for which the vast majority of requests access client-specific data structures~(e.g.,~to renew session leases~\cite{burrows06chubby}).

\section{Background \& Problem Statement}
\label{sec:problem}

Providing the agreement stage of a replicated system with information about request conflicts makes it possible to significantly increase consensus efficiency by limiting the ordering to requests that interfere with each other~\cite{pedone99generic}. In this section, we analyze existing approaches that apply this general concept. Notice that (although tackling a related problem) our discussion does not include the recently proposed ezBFT~\cite{arun19ezbft}, as since publication the protocol has been found to contain safety, liveness, and execution consistency violations~\cite{shrestha19revisiting}.

\begin{figure}
	\includegraphics{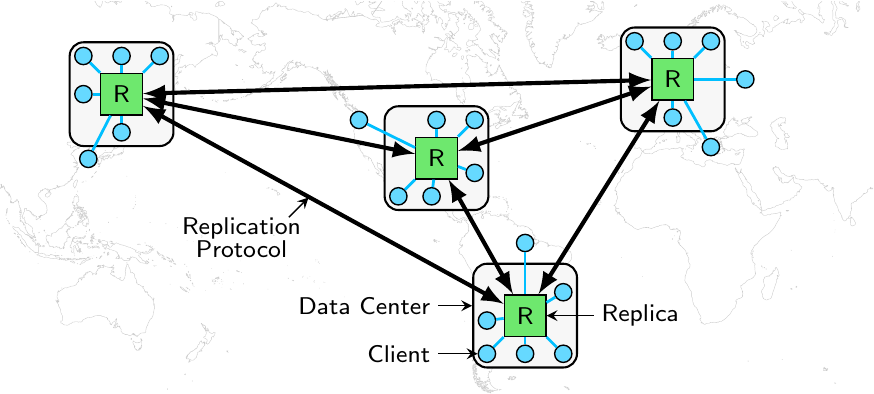}
	\vspace{-5mm}
	\caption{Geo-distributed state-machine replication}
	\label{fig:background}
\end{figure}

\subsection{Existing Approaches}
\label{sec:existing-approaches}

Based on their design goals and characteristics, existing protocols can be classified into the following three categories.

\headline{Crash Tolerance}
One of the first leaderless consensus algorithms focusing on conflicting requests was EPaxos~\cite{moraru13there}, which enables all replicas in the system to initiate the agreement process for new client requests. In geo-distributed deployments where clients are scattered across the globe~(see Figure~\ref{fig:background}), this property often significantly improves latency as each client can directly submit requests to its local replica, instead of all clients having to contact the same central leader. If a quorum of replicas agrees on a proposed request's conflicts, EPaxos allows the proposing replica to immediately commit and process the request; otherwise, the replica is required to execute a sub-protocol responsible for resolving the conflict discrepancies. Building on the same general idea, the recently proposed Atlas~\cite{enes20state} protocol offers several improvements over EPaxos, including for example the use of smaller quorums as well as the ability to commit requests early even if the conflict reports of different replicas do not match exactly~(see Section~\ref{sec:related} for details). Both EPaxos and Atlas tolerate crashes.

\headline{BFT with Additional Replicas}
The quorum-based Q/U~\cite{abd-el-malek05qu} offers resilience against Byzantine faults without the need for a global leader, however to do so it requires \mbox{$5f+1$} replicas. Byblos~\cite{bazzi21clairvoyant}, a BFT protocol tailored to permissioned ledgers, reduces the replication cost to $4f+1$~servers by determining the execution order of transactions based on a leaderless \linebreak non-skipping timestamp algorithm that is driven by clients.

\headline{Global Leader Replica}
Byzantine Generalized Paxos~\cite{pires18generalized} shows that it is possible for a BFT protocol to only order conflicting requests with a minimum of $3f+1$~replicas. However, to resolve request-conflict discrepancies between replicas the protocol resorts to a global leader which then sequentializes the affected requests. For this purpose, followers need to provide the leader with information about all requests they have previously voted for, making conflict resolution an expensive \linebreak undertaking, as confirmed by our experiments in Section~\ref{sec:evaluation}.

\subsection{Problem Statement}
\label{sec:problem-statement}

The analysis above has shown that existing approaches explore different tradeoffs with regard to fault model, replica-group size, and the existence of a global leader replica. In contrast, our goal in this paper is to integrate several desirable properties within the same state-machine replication protocol:
\begin{itemize}
	\item \textbf{Byzantine Fault Tolerance:} The protocol should tolerate up to $f$~replica faults as well as an unlimited number of faulty clients that possibly collude with faulty replicas.
	\item \textbf{Resource Efficiency:} To also support small deployments, the protocol must require a minimum of $3f+1$~replicas.
	\item \textbf{Leaderlessness:} To avoid a bottleneck and enable clients to submit requests to their nearest replica, the protocol must not rely on a single global leader replica. This should not only apply to normal-case operation, but also to the task of reconciling discrepancies between replicas.
	\item \textbf{Low Latency:} In the absence of discrepancies, the agreement process should complete within three communication steps, which is optimal for the targeted systems.
	\item \textbf{Bounded State:} To avoid an infinite accumulation of consensus state, in contrast to other leaderless protocols~(e.g.,~EPaxos), the protocol should comprise a checkpointing mechanism for garbage-collecting such state. In addition, the protocol's execution stage should also be able to operate with a bounded amount of memory when determining the request execution order based on the conflict dependencies reported by the agreement stage.
\end{itemize}
In the following, we show that it is possible to unite these properties in a single state-machine replication protocol.

\section{\system}
\label{sec:approach}

\system is a leaderless BFT protocol designed to minimize latency in wide-area settings. This section first gives an overview of \system and then provides details on different protocol mechanisms;
for pseudo code \ifextended{refer to Appendix~\ref{sec:appendix-pseudo-code}}{please refer to \cite{isos-extended}}.

\subsection{Overview}
\label{sec:architecture}

\system requires a minimum of $N=3f+1$ replicas to tolerate $f$~faults and enables each of the replicas to order client requests without the involvement of a global leader.
This allows clients to submit their requests to the nearest replica and thereby avoid lengthy detours.
When a replica receives a request from a client, the replica acts as \textit{request coordinator} and manages the replication of the request to all other replicas, which for this specific request serve as \textit{followers}. That is, to prevent bottlenecks as well as disruptions due to costly election procedures, replica roles in \system are not assigned globally as in many \linebreak other BFT protocols~\cite{distler21byzantine}, but instead on a per-request basis.

To order client requests as coordinator, each replica~$r_i$ maintains its own sequence of \textit{agreement slots} which are uniquely identified by sequence numbers~\mbox{$s_i=\langle r_i,sc_i\rangle$}, with $sc_i$ representing a local counter. Apart from its own agreement slots, each replica also stores information about other replicas' agreement slots for which the local replica acts as follower.

\headline{Consensus Fast Path}
Having received a new request, a coordinator allocates its next free agreement slot and creates a \textit{dependency set} containing all conflicts the new request has to previous requests already known to the coordinator. As illustrated in Figure~\ref{fig:overview} for request $A$, the coordinator then initiates the consensus process by forwarding the request together with the dependency set to its followers. In a next step, a coordinator-selected quorum of $2f$~followers react by computing and broadcasting their own dependency set for the request. If all of these followers report the same dependencies as the coordinator, the consensus process completes at the end of another protocol phase, that is after three communication steps; we refer to this scenario as \system's \emph{fast path}. Notice that the fast path in \system is not exclusive to non-conflicting requests, but as illustrated by the example of request~$B$ in Figure~\ref{fig:overview} can also be taken by conflicting client requests.

\begin{figure}
	\includegraphics{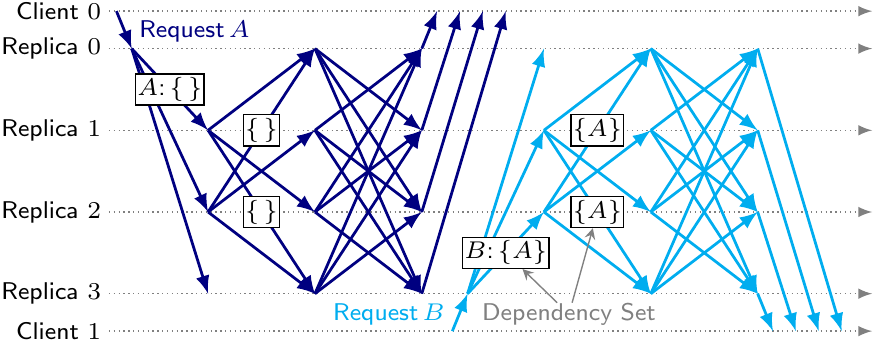}
	\caption{Fast-path ordering of two conflicting requests $A$ and $B$ in \system}
	\label{fig:overview}
	\vspace{-.4mm}
\end{figure}

\headline{Reconciliation \& View Change}
If the coordinator determines that the fast-path quorum for a request is no longer possible, it triggers \system's reconciliation mechanism which is responsible for resolving the request-conflict discrepancies between replicas by deciding on a consistent dependency set.
In case of a faulty leader or faulty followers in the coordinator-selected quorum, the replicas initiate a view change for the affected agreement slot and continue to perform reconciliation.

\headline{Request Execution}
\system replicas rely on a deterministic algorithm to determine the execution order of requests based on the dependency sets they agreed on in the consensus process. Collecting dependency sets from a quorum of replicas ensures that conflicting requests, even when proposed by different coordinators at the same time, will pick up a dependency between them and thus guarantee a consistent execution order. For non-conflicting requests, there are no dependencies to consider, meaning that a replica is allowed to independently process such a request once it has been committed by the agreement stage. After executing a request, the replicas send a reply to the client which waits for $f+1$~matching replies to ensure \linebreak that at least one of the replies originates from a correct replica.

\headline{Checkpointing}
\system relies on 
checkpointing to limit the amount of memory required by the agreement protocol and to allow replicas that have fallen behind to catch up.
To create a consistent checkpoint, all replicas have to capture a copy of the application state after executing the exact same set of requests.
As each replica can independently propose and execute requests, in contrast to traditional protocols such as PBFT~\cite{castro99practical}, in \system there are no predefined points in time~(e.g.,~specific sequence numbers) at which all replicas have the same application state.
To solve this problem, \system introduces \emph{checkpoint requests} which are agreed upon by the replicas and act as a barrier separating the requests that should \linebreak be covered by a checkpoint from the ones that should not.

\subsection{Fast Path}
\label{sec:fast-path}

When a new request~$r=\langle \msgreq, x, t, o\rangle_{\sigma_x}$ for command~$o$ from client~$x$ arrives at a replica, the replica serves as coordinator for the request; $t$ is a client-local timestamp that increases for each request and enables replicas to ignore duplicates.

\headline{DepPropose Phase}
To start the fast path, the coordinator selects its agreement slot with the lowest unused sequence number and computes the dependency set containing sequence numbers of requests that conflict with request~$r$.
For this purpose, the coordinator takes all known requests from both its own and other replicas' agreement slots into account.
Requests of the same client are automatically treated as conflicting with each other, independent of their content.
This ensures that all correct replicas will later execute the requests of a client in the same order and therefore discard the same requests as duplicates.
As a consequence, faulty clients cannot introduce inconsistencies between correct replicas by assigning the same timestamp to two non-conflicting requests.
On correct clients, on the other hand, the client-specific request dependencies have no impact as correct clients commonly only submit a new \linebreak request after having received a result for their previous one. 

To limit the size of the set, the coordinator for each replica only includes the sequence number of the latest conflicting request, thereby treating the replica's earlier requests as implicit dependencies~\cite{moraru13there}.
This approach potentially introduces~(unnecessary) additional dependencies, however it offers two major benefits: (1)~a compact dependency set in general is significantly smaller than a full set explicitly containing all conflicts would be, and (2)~since correct replicas only accept and process compact dependency sets, a faulty replica cannot slow down the agreement process by distributing huge sets.

Having assembled the dependency set~$D$ for request~$r$ in agreement slot~$s_i$, the coordinator~$co$ selects a quorum~$F$ containing the IDs of the $2f$~followers to which it has the lowest communication delay.
As shown in Figure~\ref{fig:paths}~(left), the coordinator then broadcasts a $\langle \msgpropose, s_i, co, h(r), D, F \rangle_{\sigma_{co}}$ message together with the full request to all of its follower replicas; $h(r)$~is a hash that is computed over client request~$r$.

\headline{DepVerify Phase}
Follower replicas accept a \msgpropose if the message originates from the proper coordinator and is accompanied by a client request with matching hash~$h(r)$.
A follower only sends a \msgverify in the next protocol phase if it is part of the quorum~$F$.
In such case, follower~$f_i$ calculates its own dependency set~$D_{f_i}$ for request~$r$ and broadcasts the set in a $\langle \msgverify, s_i, f_i, h(dp), D_{f_i} \rangle_{\sigma_{f_i}}$ message to all replicas, with $dp$~referring to the corresponding \msgpropose.

Followers strictly process the \msgpropose{}s of a coordinator in increasing order of their sequence numbers, thereby ensuring that a coordinator cannot skip any sequence numbers. 
Furthermore, they only compile and send the \msgverify for a \msgpropose once they know that consensus processes have been initiated for all agreement slots listed in the \msgpropose{}'s dependency set.
A follower has confirmation of the start of the consensus process if it fully processed a \msgpropose, received $f+1$~\msgverify{}s, or triggered a view change for a slot.
Waiting for the conflicting slots to begin ensures that all dependencies in the dependency set will eventually complete agreement and thus guarantees that a faulty coordinator cannot block execution of a client request by including dependencies to non-existent requests.

\headline{DepCommit Phase}
When a replica receives a \msgverify it checks that the included hash~$h(dp)$ matches the slot's \msgpropose and that the sender is part of the quorum~$F$.
As before for the \msgpropose, the replica then waits until it knows that all agreement slots contained in the dependency set~$D_{f_i}$ will finish eventually.
To continue with the fast path, a replica must complete the predicate \textit{fp-verified}, which requires a valid \msgpropose from the coordinator and matching \msgverify{}s from the $2f$~followers selected in the quorum~$F$.
The set of \msgverify{}s matches the \msgpropose if either all \msgverify{}s have the same dependency set as in the \msgpropose or if all additional dependencies are included in at least $f+1$~\msgverify{}s.
In the latter case, at least one correct replica has reported the additional dependencies, which ensures that these dependencies will be included in the fast path or the reconciliation path~(see Section~\ref{sec:slow-path}), independent of the behavior of faulty replicas.
The \msgpropose and \msgverify{}s yield a Byzantine majority quorum of $2f+1$~replicas, thereby guaranteeing that only a single proposal can complete, as correct replicas only accept the first valid \msgpropose.

Once \textit{fp-verified} holds, a replica broadcasts a corresponding $\langle \msgfpcommit, s_i, r_i, h(\vec{dv}) \rangle_{\sigma_{r_i}}$ message in which $\vec{dv}$~refers to the set of \msgverify{}s received from the followers in~$F$.
As each correct replica includes \msgverify{}s from the same followers, they all will use the same set~$\vec{dv}$ to calculate~$h(\vec{dv})$.

An agreement slot in \system is \textit{fp-committed} once a replica has obtained matching \msgfpcommit{}s from $2f+1$~replicas (possibly including itself).
At this point, the replica forwards the request to the execution~(see Section~\ref{sec:request-execution}), together with the union of the dependency sets of the \msgpropose and \msgverify{}s.
The quorum guarantees that if a request commits, then enough replicas have \textit{fp-verified} it and consequently the request will be decided by (potential) later view changes.

\begin{figure}
	\includegraphics{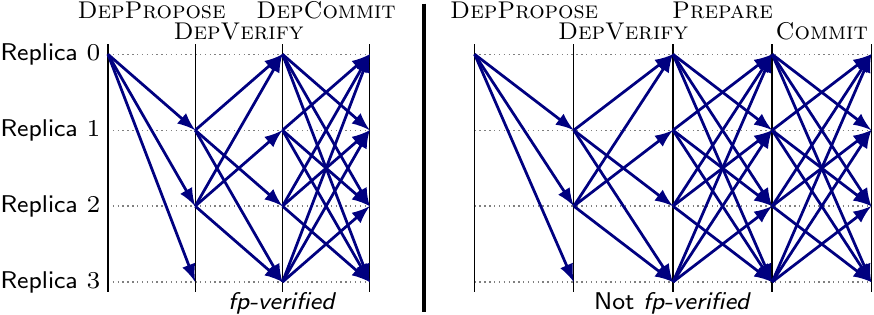}
	\vspace{-4mm}
	\caption{Fast path (left) and abandoned fast path + reconciliation (right)}
	\label{fig:paths}
\end{figure}

\subsection{Reconciliation Path}
\label{sec:slow-path}

If a replica observes that completing \textit{fp-verified} is not possible due to diverging dependency sets, the replica abandons the fast path and starts reconciliation (as illustrated on the right side of Figure~\ref{fig:paths}).
The main responsibility of \system's reconciliation mechanism is to transform the diverging dependency sets from the fast path into a single dependency set that is agreed upon by all correct replicas.
To ensure that fast path and reconciliation path cannot reach conflicting decisions regarding the dependency set, a correct replica that has reached \textit{fp-verified} (and therefore already sent a \msgfpcommit on the fast path) does not contribute to the reconciliation path.

\headline{Prepare Phase}
Upon switching to the reconciliation path, a replica stops participating in the fast path and broadcasts a $\langle\msgprepare, v_{s_i}, s_i, r_i, h(\vec{dv})\rangle_{\sigma_{r_i}}$ message in which $\vec{dv}$ is the set of previously received \msgverify{}s; $v_{s_i}$ denotes a view number, which in contrast to traditional BFT protocols~\cite{castro99practical} in \system is not global, but a variable specific to the individual agreement slot.
That is, for each request that enters reconciliation the view number starts with its initial value of~$-1$.

\headline{Commit Phase}
After a replica has obtained $2f+1$~\msgprepare{}s matching the set of known \msgverify{}s, the replica has \textit{rp-prepared} the agreement slot and continues with broadcasting a $\langle \msgcommit, v_{s_i}, s_i, r_i, h(\vec{dv})\rangle_{\sigma_{r_i}}$ message.
Having collected $2f+1$ \msgcommit{}s from different replicas with matching hash $h(\vec{dv})$, the replica has \textit{rp-committed} the request and forwards it to the execution, together with the union of the dependency sets of all \msgverify{}s and the associated \msgpropose.

\headline{Invariant}
\textit{An agreement slot in \system can either \textit{fp-commit} or \textit{rp-prepare}.}
As sending a \msgfpcommit and sending a \msgprepare are mutually exclusive, correct replicas can either collect enough \msgfpcommit{}s from a quorum to \textit{fp-commit} the fast path or enough \msgprepare{}s to \textit{rp-prepare} the reconciliation path, but never both, thus ensuring agreement among replicas.

\subsection{View Change}
\label{sec:view-change}

\vspace{-.2mm}

In case the agreement for a slot fails to complete within a predefined amount of time~(see Section~\ref{sec:progress-guarantees}), replicas in \system initiate a view change for the specific agreement slot affected.

\headline{ViewChange Phase}
Once a replica decides to abort a view, the replica stops to process requests for the old view and broadcasts a $\langle\msgvc,v_{s_i},s_i,r_i,certificate\rangle_{\sigma_{r_i}}$ message for the new view~$v_{s_i}$ to report the agreement-slot state in \linebreak the form of a $certificate$ of one of the following types:

\begin{itemize}
	\item A \textit{fast-path certificate ($FPC$)} consists of a \msgpropose message from the original coordinator and a set of $2f$~corresponding \msgverify messages from different followers matching the \msgpropose, thereby confirming that the agreement slot was \textit{fp-verified}.
	\item A \textit{reconciliation-path certificate ($RPC$)} consists of the original \msgpropose, $2f$ matching \msgverify{}s, and $2f+1$ matching \msgprepare{}s from different followers. The \msgprepare{}s must be from the same view. Together, these messages confirm the agreement slot to be \textit{rp-prepared}.
\end{itemize}

\noindent{}If available, a replica includes an $RPC$ for the highest view in its own \msgvc message, resorting to an $FPC$ as alternative.
If neither of the two certificates exists, the replica sends the \msgvc message without a certificate.

In case a replica receives $f+1$ \msgvc{}s for sequence number~$s_i$ with a view higher than its own, the replica switches to the $f+1$-highest view received for that agreement slot and broadcasts a corresponding \msgvc message.

\headline{NewView Phase}
The view change for a request is managed by a coordinator that is specific to the request's agreement slot~$s_i$. For a new view~$v_{s_i}$, the coordinator is selected as $co = (s_i.r_i + max(0, v_{s_i}))~\textrm{mod}~N$.
Having collected valid \msgvc{}s for its view from a quorum of $2f+1$~replicas, the coordinator determines the result of the view change. For this purpose, it deterministically selects a request based on the certificate with the highest priority: first $RPC$, then $FPC$.

If both a reconciliation-path certificate and a fast-path certificate exist at the same time, it is essential for the coordinator to determine the view-change result based on the re\-con\-ci\-lia\-tion-path certificate.
According to the reconciliation-path invariant, this path can only \textit{rp-prepare} if the fast path does not \textit{fp-commit}.
Thus, the fast-path certificate stems from up to $f$ replicas that tried to complete the \msgfpcommit phase but did not finish it, meaning that the certificate can be ignored.
The reconciliation path, on the other hand, might have completed and thus the view change must keep its result.
If no certificate exists, the view-change result is a no-op request with empty \linebreak dependencies, which later will be skipped during execution.

To install the new view, the coordinator broadcasts a $\langle\msgnewview,v_{s_i}, s_i, co, dp, \vec{dv}, VCS\rangle_{\sigma_{co}}$ message in which $dp$ is the \msgpropose, $\vec{dv}$ are the accompanying \msgverify{}s, and $VCS$ is the set of $2f+1$~\msgvc{}s used to determine the result.
If no certificate exists, $dv$ is replaced by a no-op request and $\vec{dv}$ is empty.
After having verified that the coordinator has correctly computed the \msgnewview, the other replicas follow the coordinator into the new view.
There, the \msgnewview's \msgpropose and \msgverify{}s are used to resume with the reconciliation path at the corresponding step~(see Section~\ref{sec:slow-path}), just for a higher view.
In case a request is replaced with a no-op during the view change, the request coordinator proposes the request for a new agreement slot.

\subsection{Progress Guarantee}
\label{sec:progress-guarantees}

In the following, we discuss several liveness-related scenarios and explain how \system handles them to ensure that requests proposed by correct replicas eventually become executable.

\headline{Fast Path}
Faulty replicas in \system may try to prevent correct replicas from making progress by not properly participating in the consensus process. For example, a faulty replica~$r_i$ may send a \msgpropose for a sequence number~$s_i$, but only to one correct replica~$r_j$ and not the others. Replica~$r_j$ thus must include sequence number~$s_i$ as dependency in its own future proposals, meaning that other replicas can only process $r_j$'s proposals if they also know about~$s_i$. To ensure that the system in such case eventually makes progress despite replica~$r_i$'s refusal to properly start the consensus for~$s_i$, correct followers in \system start a \textit{propose timer} with a timeout of $2\Delta$ whenever they receive a \msgpropose; $\Delta$ is the maximum one-way delay between replicas~(see Section~\ref{sec:model}). If the propose timer expires or a view change is triggered and the follower has not collected $2f$~matching \msgverify{}s in the meantime, the follower broadcasts the affected \msgpropose~(which does not include the full client request, see Section~\ref{sec:fast-path}) to all other follower replicas, thereby enabling them to move on.

\headline{Agreement}
To monitor the agreement progress of a slot, replicas in \system start a \textit{commit timer} with a timeout of $9\Delta$ once they know that the consensus process for a slot has been initiated. This is the case if a replica has (1)~sent its \msgpropose, (2)~(directly or indirectly) received a valid \msgpropose and learned that its dependencies exist or (3)~obtained $f+1$ \msgverify messages proving that at least one correct replica has accepted a \msgpropose for this slot.
If the commit timer expires, a replica triggers a view change.
\ifextended{The value of the commit timeout is explained in Appendix~\ref{sec:appendix-pseudo-code}.}{Please refer to \cite{isos-extended} for an explanation of the timeout value.}

Forwarding the \msgpropose after the propose timer expires (see above) and listening for \msgverify messages ensures that every correct replica will eventually learn that a proposal for the agreement slot exists and thus start the commit timer.
This in turn guarantees that either $f+1$~correct replicas commit a client request or trigger a view change.

\headline{Recovering the Fast-Path Quorum}
If the quorum~$F$ proposed by a fast-path request coordinator includes faulty replicas, it is possible that these replicas do not send \msgverify messages and thus prevent requests from being ordered in the agreement slot.
In such case, after the agreement slot was completed with a no-op by a view change,  the request coordinator selects a different set of $2f$~followers and proposes the request for a new agreement slot.
This ensures that eventually all replicas in quorum~$F$ are correct which allows the agreement to complete.

\headline{Lagging Replicas}
As the active involvement of \mbox{$2f+1$}~replicas is sufficient to commit a request in \system, there can be up to $f$~correct but lagging replicas that do not directly learn the outcome of a completed agreement process. Furthermore, as the agreement processes of different coordinators advance largely independent of each other, different replicas may lag with respect to different coordinators. To resolve circular-waiting scenarios under such conditions, an \system replica can query others for committed requests. If $f+1$~replicas~(i.e.,~at least one correct replica) report a request to have committed for an agreement slot, the lagging replica also regards the request as committed. Since $2f+1$~replicas are required to complete consensus, for each completed slot there are at least $f+1$~correct\linebreak  replicas that can assist lagging replicas in making progress.

\headline{Crashed Replicas}
If a coordinator crashes, the effects of the crash are limited to the slots the coordinator has started prior to its failure. Once these slots have been completed~(if necessary through view changes), there are no further impairments as the failed coordinator does no longer propose new requests, and thus there are no new dependencies on the coordinator.

\subsection{Request Execution}
\label{sec:request-execution}

Using committed requests and their dependency sets as input, the execution stage of a replica is responsible for determining the order in which the replica needs to process these requests. For correct replicas to remain consistent with each other, they all must execute conflicting requests in the same relative order. Non-conflicting requests on the other hand may \linebreak be processed by different replicas at different points in time. In the following, we explain how \system ensures that these requirements are met even if faulty replicas manipulate dependencies.

\begin{figure}
	\includegraphics{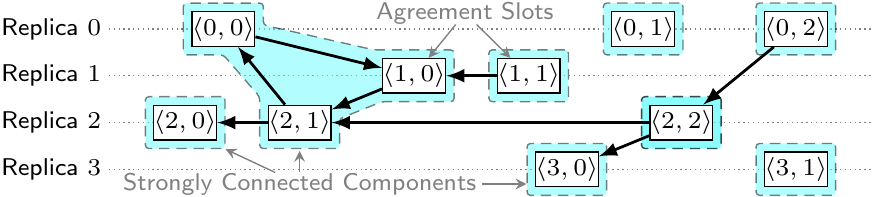}
	\vspace{-4mm}
	\caption{Strongly connected components in an execution dependency graph}
	\label{fig:execution}
\end{figure}

\headline{Regular Request Execution}
\label{sec:exec-algorithm}
For each committed request, the execution builds a dependency graph whose nodes are not yet executed requests which are connected by directed edges as specified in the requests' dependency sets.
This graph is constructed by recursively expanding the dependencies of the request.
If a dependency refers to  a not yet committed agreement slot, the graph expansion waits until the dependency is committed.
The execution then calculates the strongly connected components in the dependency graph and executes them in inverse topological order.
As illustrated in Figure~\ref{fig:execution}, each strongly connected component represents either a single request or  multiple requests connected by cyclic dependencies.
The inverse topological order ensures that dependencies of all requests in a strongly connected component are executed first.
For each such component, the requests are sorted and then executed according to their slot sequence number to ensure an identical execution order on all replicas.
The execution uses the timestamp in a client request to filter out duplicates.

\headline{Handling Dependency Chains}
\label{sec:exec-livelocks}
As the dependency collection for a request is a two-step process~(see Section~\ref{sec:fast-path}), it is possible that \msgverify messages include dependencies to agreement slots that were proposed after the request itself.
These slots in turn can also collect dependencies to additional future slots resulting in a temporary execution livelock~\cite{moraru13there} that delays the execution of a request until all its dependencies are committed.
Such dependency chains can either arise naturally when processing large amounts of conflicting requests~\cite{moraru13there} or due to faulty replicas manipulating dependency sets by including dependencies to future requests in their \msgverify{}s.

To handle this kind of dependency chains with a bounded amount of memory, \system replicas limit how many requests are expanded.
Specifically, for each coordinator the execution only processes a window of $k$~agreement slots.
The start of each window points to the oldest agreement slot of the coordinator with a not yet executed request.
Dependencies to requests beyond this expansion limit are treated as missing and block the execution of a request.
This bounds the effective size of dependency chains, while still allowing the out-of-order execution of non-conflicting requests within the window.

To unblock request execution, replicas use the following algorithm:
First, a replica tries to normally execute all committed requests within the execution window.
Then, for each coordinator the replica constructs the dependency graph of the oldest not yet executed request, called \textit{root node}, and checks whether its execution is only blocked by missing requests beyond the execution limit.
If this is the case, the replica ignores dependencies to the latter requests and starts execution.
However, it only processes the first strongly connected component and then switches back to regular request execution.

The intuition behind the algorithm is that the execution of root nodes occurs when only requests beyond the expansion limit are still missing~(i.e.,~at a time when all replicas see the same dependency graph).
For dependencies to other nodes \system's compact dependency representation~(see Section~\ref{sec:fast-path}) automatically includes a dependency on the root node for the associated coordinator, this ensures that dependent root nodes are executed in the same order on correct replicas.

\subsection{Checkpointing}
\label{sec:checkpointing}

The checkpoints of correct replicas in BFT systems must cover the same requests in order to be safely verifiable by comparison~\cite{eischer19deterministic}.
Traditional BFT protocols~\cite{castro99practical,distler16resource,eischer19scalable} ensure this by requiring replicas to snapshot the application state in statically defined sequence-number intervals.
In \system, this approach is not directly applicable because instead of one single global sequence of requests, there are multiple sequences~(i.e.,~one per coordinator) that potentially advance at different speeds.
To nevertheless guarantee consistent checkpoints, \system replicas rely on dedicated \textit{checkpoint requests} to dynamically determine the points in time at which to create a snapshot.
As illustrated in Figure~\ref{fig:checkpoints}, a checkpoint request conflicts with every other request and therefore acts as a barrier such that each regular client request on all correct replicas is either executed before or after the checkpoint request.

\headline{Basic Approach}
A checkpoint request in \system is a special empty request that is known to all replicas and when processed by the execution triggers the creation of a checkpoint.
Each correct replica is required to propose the checkpoint request for every own agreement slot with sequence number $sc_i~\textrm{mod}~cp\_interval = 0$; $cp\_interval$ is a configurable constant that also defines the minimum size of the agreement ordering window~(i.e.,~$2*cp\_interval$), that is the number of slots per coordinator for which a replica needs information.

Relying on a checkpoint request to determine when to create a snapshot in \system has the key benefit that replicas, as a by-product of the consensus process for this request, also automatically agree on the client requests the checkpoint must cover. Specifically, based on the checkpoint request's dependency set replicas know exactly which client requests they are \linebreak required to execute prior to taking the application snapshot.

Having created the checkpoint, a replica broadcasts a $\langle \msgcp, cp.seq, r_i, barrier, h(cp)\rangle_{\sigma_{r_i}}$ message to all other replicas; $cp.seq$ is a monotonically increasing checkpoint counter, $barrier$ refers to the requests included in the checkpoint~(i.e.,~the dependency set plus the checkpoint request itself), and~$h(cp)$ represents a hash of the checkpoint content.

Once a replica has collected $2f+1$ matching checkpoint messages from different replicas, the messages form a checkpoint certificate that proves the stability of the checkpoint. After obtaining such a certificate, a replica can garbage-collect all earlier state covered by the checkpoint, including requests kept for conflict calculations.
As a substitute, a replica from this point on uses $barrier$ as minimum dependency set.

\begin{figure}
	\includegraphics{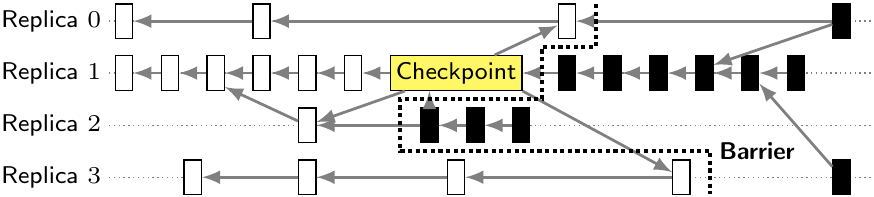}
	\vspace{-4mm}
	\caption{Checkpoint request serving as barrier for regular client requests}
	\label{fig:checkpoints}
\end{figure}

\headline{Checkpoint-specific View Change}
While regular agreement slots may eventually result in a no-op being committed~(see Section~\ref{sec:view-change}), \system guarantees that the proposal of a checkpoint request will eventually succeed within its original checkpoint slot. Our solution to achieve this relies on an auxiliary \msgverify that a replica additionally includes in its \msgvc when starting a view change for a checkpoint slot. The auxiliary \msgverify contains a placeholder hash as well as a dependency set for the checkpoint request. If the replica has previously participated in the fast path, the dependency set is identical with the one from the replica's own \msgpropose or \msgverify, otherwise the replica computes a new dependency set for the checkpoint request. Notice that due to the fact that the content and sequence numbers of checkpoint requests are known in advance, a replica is able to create such an auxiliary \msgverify even if it has not received the actual \msgpropose for the checkpoint slot.

Utilizing the auxiliary \msgverify{}s, we are able to extend the certificate list of Section~\ref{sec:view-change} with a third option: a \emph{checkpoint request certificate~($CRC$)} that is selected if neither of the two other certificates is available. The $CRC$ consists of $2f+1$ auxiliary \msgverify{}s verified to only include known dependencies and can be used in the new view to agree on a common dependency set. Since for checkpoint slots, the $CRC$ is always available as a fallback, there is no need for a view-change coordinator to introduce a no-op request.

\headline{Checkpoint-specific Execution}
Being generally treated like regular client requests, a checkpoint request can be part of a dependency cycle in which some requests should be processed before the checkpoint, while others are to be executed afterwards.
To handle such a scenario, \system's execution processes strongly connected components in a special way if they contain checkpoints.
First, it merges the dependency sets of all checkpoint requests included in a strongly connected component, adding the checkpoint requests themselves to the merged set.
Next, the merged set is bounded to not exceed the expansion limit described in Section~\ref{sec:exec-livelocks}, and to include all requests before the first not yet executed request of each replica.
The resulting set now acts as a barrier defining which requests should be covered by the checkpoint and which should not.
In the final step, the execution uses the barrier to only execute client requests before the barrier, followed by the merged checkpoint request.
For the remaining requests after the barrier, a new dependency graph is constructed and used to order requests.
Restarting the execution algorithm for these requests ensures that they are executed the same way as a lagging \linebreak replica would do if it applied the checkpoint to catch up.

\subsection{Correctness (Proof sketch; full proof in Appendix~\ref{sec:appendix-pseudo-code})}
\label{sec:dependency-correctness}

\headline{Safety}
\textit{All correct replicas that commit a slot must decide on the same request and dependencies.}
A correct replica can only commit on the fast or reconciliation path if it has collected a quorum of \msgverify{}s or \msgprepare{}s, which ensures that all replicas agree on the same request.
As shown in Section~\ref{sec:slow-path}, committing the fast or reconciliation path is mutually exclusive, meaning that within a view all replicas arrive at the same result.
The final dependencies for a slot are defined by the \msgpropose and the set $\vec{dv}$ of $2f$ \msgverify{}s whose hash $h(\vec{dv})$ is included in the \msgfpcommit and \msgcommit messages, respectively. This ensures that all replicas agree on the dependencies.
After a successful commit, at least $f+1$~correct replicas have collected a certificate for the fast or reconciliation path, and thus the certificate will be included in future view changes.

\headline{Execution Consistency {\normalfont (as used in EPaxos~\cite{moraru13there})}}
\textit{If two con\-flic\-ting requests \reqa and \reqb are committed, all replicas will execute them in the same order.}
This is achieved by ensuring that the two requests are connected by a dependency such that either \reqa depends on \reqb, or \reqb depends on \reqa, or both depend on each other.
All three cases result in the execution consistently ordering the requests before processing them.
The dependencies for a request are collected from a quorum of $2f+1$~replicas using \msgpropose and \msgverify messages.
If requests \reqa and \reqb are proposed by different replicas at the same time, their dependency collection quorums will overlap in at least $f+1$ replicas, of which at least one replica must be correct.
This replica will either receive \reqa or \reqb first and thus add a dependency between them.
Therefore, two conflicting requests are always connected by a dependency.

Note that a malicious coordinator proposing different \msgpropose{}s to its followers cannot cause missing dependencies.
Either the same \msgpropose is fully processed by at least $f+1$~correct replicas (which ensures dependency correctness), the faulty \msgpropose{}s are ignored, or none of the \msgpropose{}s gathers $2f$~\msgverify{}s, thus causing the slot to be filled with a no-op during the following view change.
In the latter case, no dependencies from or to the slot are necessary, as a no-op command does not conflict with any other request.

The dependency cannot be lost when switching between protocol paths or during a view change.
The reconciliation path carries over the dependency sets from the fast path and cannot introduce new dependencies in the agreement process.
Replicas that learn about a client request in a view change have no influence on the dependency calculations for the request.

\headline{Invariant}
\textit{The view change either selects the (only) request that was \textit{fp-verified} or \textit{rp-prepared}, or a no-op.}
We proof this by induction.
Only a single \msgpropose can collect $2f$ matching \msgverify{}s in a slot.
Thus, no fast or reconciliation path certificate can exist for any other request, as constructing a certificate requires a matching set of \msgverify{}s from a quorum of replicas.
The view change only selects a request with a certificate or a no-op, and hence all future re\-con\-ci\-lia\-tion-path executions can only decide one of the two.
This guarantees that a slot either commits the request initially sent to a majority of replicas or a (by definition) non-conflicting no-op.
Requests that were not properly proposed to a quorum of replicas will therefore be replaced with a no-op. This \linebreak ensures that all ordered requests have proper dependency sets.

\section{Evaluation}
\label{sec:evaluation}

\def\csp{CSP\xspace}
\def\pbft{PBFT\xspace}

In this section, we experimentally evaluate \system together with other protocols in a geo-replicated setting. For a fair comparison, we focus on BFT protocols and implement them in a single codebase written in Java:
(1)~\textbf{\pbft}~\cite{castro99practical} represents a protocol that pursues the traditional concept of relying on a central global leader replica to manage consensus.
(2)~\textbf{\csp}, short for Centralized Slow Path, refers to a hybrid approach which, similar to Byzantine Generalized Paxos~(BGP)~\cite{pires18generalized}, combines a leaderless fast path with a leader-based slow path for conflict resolution.
We decided to create \csp because BGP requires its leader replica to share large sets of previously ordered requests to resolve conflicts, which in practical use-case scenarios results in unacceptable overhead.
Since \csp's slow path does not suffer from this problem, we expect \csp's results to represent a best-case approximation of BGP's performance.\linebreak
(3)~\textbf{\system} in contrast to the other two protocols is entirely leaderless, in both the fast path as well as during reconciliation.

We conduct our experiments hosting the replicas in virtual machines (t3.small, 2 VCPUs, 2GB RAM, Ubuntu 18.04.5 LTS, OpenJDK 11) in the Amazon EC2 regions in Oregon, Ireland, Mumbai, and Sydney.
Our clients run in a separate virtual machine in each region. \csp's slow-path leader resides in Oregon.
All messages exchanged between replicas are signed with 1024-bit RSA signatures.
As the communication times between replicas vary between 59 and 127\,ms, we set $\Delta$ to $200$\,ms.
Replicas use $cp\_interval=2$,$000$ to create new checkpoints and an expansion limit of $20$ for the request execution.
Each coordinator accumulates new client requests in batches of up to 5 requests before proposing them for ordering.

As application for our benchmarks, we use a key-value store for which clients issue read and write requests in a closed loop.
Write requests modifying the same key conflict with each other.
In contrast, read requests for a key only conflict with write requests but not with other read requests.

\subsection{Latency}

In our first experiment, we use a micro benchmark~(200 bytes request payload, 10~clients per region) to compare the response times experienced by clients in the three systems.
To control the rate of requests that can conflict with each other, we follow the setup of EPaxos~\cite{moraru13there} and ATLAS~\cite{enes20state} and let clients issue write requests for a fixed key with a probability~$p$, and for a unique key otherwise.
We use conflict rates of 0\%, 2\%, and 5\% to evaluate typical application scenarios of \system; for comparison, EPaxos considers low conflict rates between 0\% and 2\% as most realistic~\cite{moraru13there}.
In addition, to present the full picture we repeat our experiment with conflict rates of 10\% and 100\% for completeness.
For \pbft, which is not affected by the conflict rate, we instead measure the latency for each possible leader location.
The results of this experiment are presented in Figure~\ref{fig:micro-locations}.
For clarity, we omit the \csp numbers for low conflict rates of 0\% and 2\% as they are dominated by the fast \linebreak path and thus similar to the corresponding results of \system.

\begin{figure}
	\includegraphics{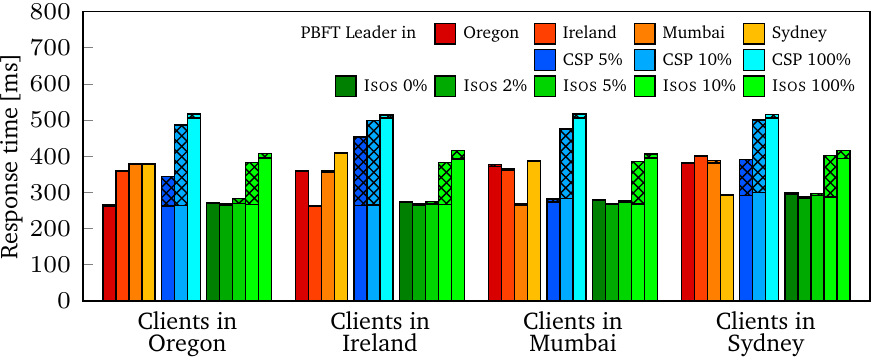}
	\vspace{-4mm}
	\caption{50th~(\raisebox{-.3mm}{\protect\tikz \protect\node[draw, minimum width=7, minimum height=7] {};}) and 90th~(\raisebox{-.3mm}{\protect\tikz \protect\node[draw, minimum width=7, minimum height=7, postaction={pattern=crosshatch, pattern color=black}] {};}) percentiles of response times for clients at different geographic locations, issuing requests with various conflict rates.}
	\label{fig:micro-locations}
\end{figure}

In \pbft, the median response times for clients in a region heavily depend on the current location of the leader replica.
For clients in Ireland, for example, the response times can increase by up to 56\% when the leader replica is not located in Ireland but in a different region.
This puts all clients at a disadvantage whose location differs from that of the leader.
In contrast, for typical low conflict rates of 2\%, \system in each region achieves median and 90th percentile response times similar to those of the best \pbft configuration for that region.
However, \pbft due to its reliance on a single leader replica can only provide optimal response times for a single region at a time,
whereas \system's leaderless design enables clients to submit their requests to a nearby replica and thus provides optimal response times for clients in all regions at once.

For conflict rates of 5\% and higher, the median and 90th percentile response times for \csp rise up to 517\,ms,
which is a result of the additional communication step required by the central leader to initiate the agreement on conflicting dependencies.
For comparison, the response times of \system are significantly lower even for a conflict rate of 100\% where most requests are ordered via the reconciliation path.
This illustrates the benefits of \system's design choice to refrain from a global leader, not only on the fast path but also during reconciliation.

\subsection{Throughput}
In our second experiment, we assess the relation between throughput and response times for up to 1,000 evenly distributed clients and different request sizes~(see Figure~\ref{fig:micro-throughput-small}).
For \pbft and requests with 200 bytes payload,
the average response time stays below 369\,ms for up to 400 clients and starts to rise afterwards.
The throughput reaches nearly 1,875 requests per second at which point it is limited by the leader replica saturating its CPU.
For low conflict rates of 0\% and 2\% \system, on the other hand, achieves response times below 304\,ms for up to 400 clients and reaches a throughput of up to 2,079 requests per second.
This represents an improvement of 18\% lower latency and 11\% higher throughput over \pbft, showing the benefit of clients being able to submit their requests to a nearby replica instead of forwarding it to a central leader.
Comparing \csp and \system for conflict rates above 5\% shows that \csp provides higher response times and thus lower throughput than \system,
which  is a consequence of the additional communication step necessary to initiate \csp's slow path.

Issuing large requests with 16\,kB payload from up to 600 clients, we observe that \pbft reaches a maximum throughput between 632 and 764 requests per second depending on the leader location.
At this point, the network connection of the leader, which has to distribute the requests to all other replicas, is saturated and prevents further throughput increases.
In contrast, \system reaches a maximum throughput of 1,328~requests per second, outperforming \pbft by up to 110\%.
The throughput advantage even holds for conflict rates as high as 10\%.
\system benefits from its leaderless design in which all replicas share the load of distributing requests, allowing it to \linebreak handle larger requests than a protocol using a single leader.

\begin{figure}
	\includegraphics{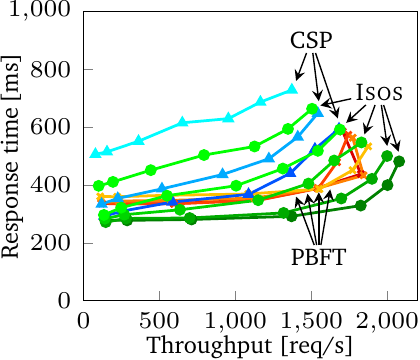}
	\hfill
	\includegraphics{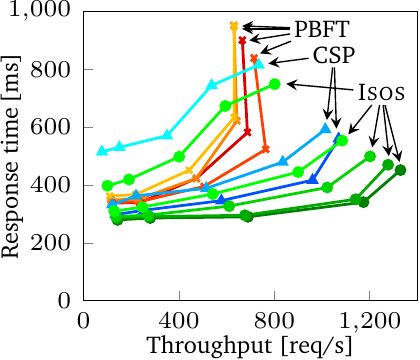}\\
	\vspace{-4mm}
	\caption{Relation between average throughput and response time for client requests with different payload sizes of 200\,bytes~(left) and 16\,kB~(right).}
	\label{fig:micro-throughput-small}
	\label{fig:micro-throughput-big}
\end{figure}

\subsection{YCSB}
In our third experiment, we run the YCSB benchmark~\cite{cooper10benchmarking} with a total of 200~clients that are evenly distributed across all regions and issue a mix of reads and writes.
The database is loaded with 1,000 entries of 1kB size.
The key accessed by a client request is selected according to the Zipfian distribution which skews access towards a few frequently accessed elements and is parameterized using the standard YCSB settings.

Figure~\ref{fig:ycsb-eval} shows the throughput achieved for different shares of read and write requests.
For the write heavy 50/50 benchmark, \system and \pbft achieve similar average throughputs of nearly 600 requests per second.
Consistent with the previous benchmarks, the throughput of \csp stays below that of \system.
For the 95/5 and 100/0 workloads, \system outperforms \pbft by 17\% and 20\%, respectively.
Due to a high fraction of read requests, these workloads have a low conflict rate, thereby allowing \system to take full advantage of its leaderless design.

\begin{figure}[b!]
	\vspace{-1mm}
	\includegraphics{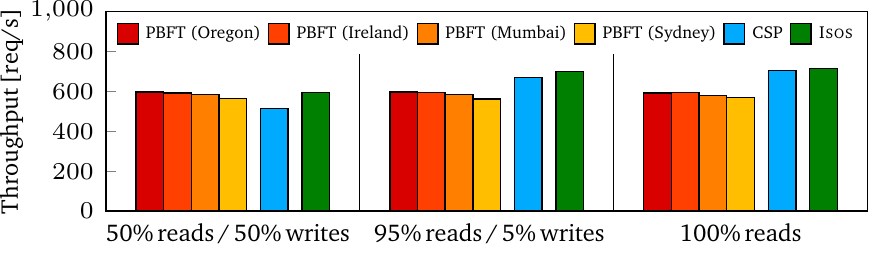}
	\vspace{-6mm}
	\caption{Average throughput for different read-write ratios in YCSB.}
	\label{fig:ycsb-eval}
\end{figure}

\section{Related Work}
\label{sec:related}
  \headline{Optimized Leader Placement}
One method to reduce the response time for systems with a central leader replica is to optimize its placement.
Archer~\cite{eischer18latency} uses clients to send probes through the agreement protocol to measure latency and thus enable the system to select a leader offering low latency.
In AWARE~\cite{berger19resilient}, replicas measure the communication latency between themselves and use the outcome to adjust replica voting weights to prefer the fastest replicas.
In \system, these approaches could be used to select optimal fast-path quorums.

\headline{Concurrent Consensus}
To distribute the work of a leader, it is possible to partition a global sequence number space onto multiple leader replicas.
Protocols like BFT-Mencius~\cite{milosevic13bounded}, Mir-BFT~\cite{stathakopoulou21mirbft}, Omada~\cite{eischer19scalable} and RCC~\cite{gupta21rcc} then run multiple ordering instances in parallel and merge them according to their sequence numbers.
In comparison to \system these protocols primarily focus on throughput and either have to wait for ordered requests from all replicas or require additional coordination  to handle imbalanced workloads.

\headline{Leaderless Consensus}
DBFT~\cite{crain18dbft} avoids using a central leader by letting replicas distribute their proposals using a reliable Byzantine broadcast and then reaching agreement on which replicas contributed proposals.
This requires at least four communication steps compared to the three of \system's fast path, resulting in higher latency.
The eventually consistent PnyxDB~\cite{bonniot20pnyxdb} uses conditional endorsements based on conflicts between requests.
An endorsement for a request becomes invalid if a conflicting request could be committed before the request, causing some requests to be dropped eventually.

\headline{Crash Faults}
PePaxos~\cite{ceolin20parallel} is a recent variant of EPaxos~\cite{moraru13there} which during execution uses the agreement's dependency sets to schedule independent strongly-connected components for parallel execution. This approach can also be integrated in \system.
Atlas~\cite{enes20state} uses a fast path based on a preselected quorum of replicas, allowing it to optimize the reconciliation of differing dependency sets.
Dependencies for an agreement slot proposed by at least $f$ replicas can be agreed on via the fast path, allowing Atlas to always take the fast path for $f=1$.
\system uses a similar optimization for its fast path requiring $f+1$ replicas to report dependencies to handle Byzantine faults.

\section{Conclusion}
\label{sec:conclusion}

\system is a fully leaderless BFT protocol for geo-replicated environments. It requires only $3f+1$~replicas and offers a fast path that orders client requests in three communication steps if request conflicts are reported by at least $f+1$~replicas.

{
	\vspace{.4mm}
	\small
	\noindent{}\emph{Acknowledgments:} This work was partially supported by the German\vspace{-.7mm}\\Research Council (DFG) under grant no. DI 2097/1-2~(``REFIT'').
}

\bibliographystyle{IEEEtran}
\bibliography{bib/paper-tiny}

\clearpage
\appendix

\subsection{\system Agreement Protocol}
\label{sec:appendix-pseudo-code}

\vspace{1mm}\hrule\vspace{-1mm}
\begin{lstlisting}[name=pseudocode]
%\pseudocode{\textbf{Variables at each replica}}%:
$p[s_j]$%\hfill%/* %\msgpropose%for agreement slot $s_j$ includes%\linebreak\hfill%fast-path quorum $F$%\,%*/
$pr[s_j]$%\hfill%/* %\msgreq%for $\msgpropose$ of agreement slot $s_j$%\,%*/
$v[s_j][f_i]$%\hfill%/* %\msgverify%for slot $s_j$ from follower $f_i$%\,%*/
$step[s_j] \in$ {init,proposed,fp-verified,fp-committed,%\linebreak%rp-verified,rp-prepared,rp-committed,view-change}
$view[s_j]$%\hfill%/* View number for slot $s_j$, initially $view[s_j] := -1$%\,%*/
$views[s_j][r_i]$%\hfill%/* Highest view number for slot $s_j$ seen for replica $r_i$%\,%*/
$cert[s_j]$%\hfill%/* Latest own certificate for slot $s_j$%\,%*/
$\Delta_{propose} := 2 \Delta$; $\Delta_{commit} := 9 \Delta$; $\Delta_{vc} := 3 \Delta$; $\Delta_{vc-commit} := 3 \Delta$; $\Delta_{query-exec} := 4 \Delta$%\vspace{2mm}\hrule\vspace{2mm}\begin{center}\textbf{Fast Path}\end{center}\hrule\vspace{2mm}%
%\pseudocode{\textbf{Request coordinator} }%$co$%\pseudocode{ \textbf{receives new} }%$r := \langle \msgreq{}, x, t, o \rangle$:
	assert $r$ correctly signed
	$s_j := \langle co, sc_j\rangle$%\hfill%/* Smallest free slot%\,%*/
	$D :=$ conflicts($r$)
	$F :=$ %\pseudocode{Quorum of}% $2f$ %\pseudocode{followers}%
	$dp := \langle \langle \msgpropose{}, s_j, co, h(r), D, F\rangle _{\sigma_{co}}, r \rangle$
	$\langle p[s_j], pr[s_j] \rangle := dp$%\label{code:agr-psj-coord}%
	$step[s_j] :=$ proposed
	%\pseudocode{Broadcast }%$d$%\pseudocode{ to all replicas}%
	%\pseudocode{Start commit timeout}% $\Delta_{commit}$ %\pseudocode{for slot}% $s_j$%\smallskip%
%\pseudocode{\textbf{Follower} }%$f_i$%\pseudocode{ \textbf{receives} }%$dp := \langle \langle \msgpropose{}, s_j, co, h(r), D, F\rangle, r \rangle$:
	%\lstbtt{pre:}% $step[s_j] =$ init
	assert $F$ is a valid fast-path quorum
	assert $pr[s_j] = \emptyset\,\wedge\,s_j.co = co$%\hfill%/*%\,%First%\,%propose%\,%from%\,%coordinator%\,%*/
	wait($D \cup s_{j-1}$)%\hfill%/* $s_{j-1}$ is the previous slot from coordinator $co$%\,%*/%\label{code:agr-wait-no-gaps}%
	%\pseudocode{If}% $p[s_j] = \emptyset$:
		%\pseudocode{Start commit timeout}% $\Delta_{commit}$ %\pseudocode{for slot}% $s_j$
		%\pseudocode{Start propose timeout}% $\Delta_{propose}$ %\pseudocode{for slot}% $s_j$
	$p[s_j] := dp.\msgpropose$
	%\pseudocode{If}% $r \neq \bot$:
		assert $r$ correctly signed
		$D_{f_i} :=$ conflicts($r$)
		$pr[s_j] := r$%\label{code:agr-psj-follower}%
		$step[s_j] :=$ proposed
		%\pseudocode{If }%$f_i \in F$:
			%\pseudocode{Broadcast }%$\langle \msgverify{}, s_j, f_i, h(dp), D_{f_i}\rangle _{\sigma_{f_i}}$%\label{code:agr-send-verify}%%\smallskip%
%\pseudocode{\textbf{Replica} }%$r_i$%\pseudocode{ \textbf{receives} }%$m := \langle \msgverify{},s_j, f_i, h(dp), D_{f_i}\rangle$:
	%\lstbtt{pre:}% $step[s_j] =$ proposed $\wedge$ $h(p[s_j]) = h(dp)$
	assert $v[s_j][f_i] = \emptyset$%\hfill%/* First verify from follower%\,%*/
	assert $f_i  \in m[s_j].F$%\hfill%/* Follower is in fast-path quorum%\,%*/
	wait($D_{f_i}$)%\label{code:agr-wait-verify}%
	$v[s_j][f_i] := m$
	$\vec{dv} := \{v[s_j][f_i]\,|\,\forall f_i \in p[s_j].F\}$
	%\pseudocode{If }%$|\,\vec{dv}\,| = 2f$:
		%\pseudocode{Stop propose timeout}% $\Delta_{propose}$ %\pseudocode{for slot}% $s_j$%\label{code:agr-stop-propose-timeout}%
		$D :=\,\cup D_{f_i} \in \vec{dv}$
		/* Every dependency is reported by at least $f+1$ followers%\,%*/
		%\pseudocode{If }%$\{d\in D\,|\,|\{f_i\,|\,\forall f_i: d \in v[s_j][f_i].D\}|\geq f+1 \} = D$:%\label{code:agr-f1-deps}%
			/* Slot $s_j$ is now fp-verified at replica $r_i$%\,%*/
			$step[s_j] :=$ fp-verified
			%\pseudocode{Broadcast }%$\langle \msgfpcommit{}, s_j, r_i, h(\vec{dv}) \rangle _{\sigma_{r_i}}$%\label{code:agr-send-fpcommit}%
		%\pseudocode{Else}%:
			%\pseudocode{Enter reconciliation path, stop participating in fast path}\label{code:agr-jump-rp}\smallskip%
%\pseudocode{\textbf{Replica} }%$r_i$%\pseudocode{ \textbf{receives} }%$\langle \msgverify{},s_j, *, *, *\rangle$%\pseudocode{ \textbf{from} }%$f+1$%\pseudocode{ \textbf{replicas}}%:
	%\pseudocode{Start commit timeout}% $\Delta_{commit}$ %\pseudocode{for slot}% $s_j$%\smallskip%
%\pseudocode{\textbf{Replica} }%$r_i$%\pseudocode{ \textbf{receives} }%$\langle \msgfpcommit{}, s_j, *, h(\vec{dv}) \rangle$%\linebreak\pseudocode{ \textbf{with identical} $h(\vec{dv})$ \textbf{from} }%$2f+1$%\pseudocode{ \textbf{replicas}}%:
	%\lstbtt{pre:}% $step[s_j] =$ fp-verified $\wedge$ $h(\{v[s_j][f_i]\,|\,\forall f_i\}) = h(\vec{dv})$
	%\pseudocode{Stop propose/commit timeout}% $\Delta_{propose}$%\pseudocode{ and }%$\Delta_{commit}$ %\pseudocode{for slot}% $s_j$
	$exec[s_j] := \langle pr[s_j], \cup D_{f_i} \in \vec{dv} \rangle$
	%\pseudocode{Forward }%$\langle pr[s_j], D, s_j \rangle$%\pseudocode{ to execution, with }%$D :=\,\cup D_{f_i} \in \vec{dv}$%\label{code:agr-fp-exec}\smallskip%
%\textbf{\lstbtt{wait}($D$):}%%\label{code:agr-wait}%
	%\pseudocode{For}% $d \in D$:
		%\pseudocode{Wait until either}%%\label{code:agr-wait-start}%
			$p[d] \neq \emptyset$ /* received a valid %\msgpropose{}\,%*/
			%\pseudocode{received $f+1$ correctly signed }\msgverify%
			%\pseudocode{received $f+1$ correctly signed }\msgvc%%\label{code:agr-wait-end}%
%\textbf{\lstbtt{conflicts}($r$):}%
	%\pseudocode{Return }%$\{s_i |\forall s_i, pr[s_i] \neq \emptyset: conflict(pr[s_i], r) \}$%\vspace{2mm}\hrule\begin{center}\textbf{Reconciliation Path}\end{center}\hrule\vspace{2mm}%
%\textbf{Timeout }%$\Delta_{propose}$%\textbf{ for slot }%$s_j$%\textbf{ expires:}%
	%\pseudocode{Broadcast }$\langle p[s_j], \bot \rangle$\pseudocode{ to all replicas}%
%\textbf{Timeout }%$\Delta_{commit}$%\textbf{ for slot }%$s_j$%\textbf{ expires:}%
	%\pseudocode{Move to new view }%$v_{s_j}+1$
%\pseudocode{\textbf{Enter reconciliation path for slot $s_j$ at replica} }%$r_i$:%\label{code:agr-enter-rp}%
	$step[s_j] :=$ rp-verified
	$\vec{dv} := \{v[s_j][f_i]\,|\,\forall f_i\}$
	%\pseudocode{Broadcast }%$\langle \msgprepare{}, view[s_j], s_j, r_i, h(\vec{dv}) \rangle _{\sigma_{r_i}}$
%\pseudocode{\textbf{Replica }%$r_i$%\pseudocode{ \textbf{receives} }%$\langle \msgprepare{}, v_{s_j}, s_j, *, h(\vec{dv}) \rangle $%\linebreak\pseudocode{ \textbf{with identical} $h(\vec{dv})$ \textbf{from} }%$2f+1$%\pseudocode{ \textbf{replicas}}\label{code:agr-2f1-prepares}%:
	%\lstbtt{pre:}% $step[s_j] =$ rp-verified $\wedge$ $view[s_j] = v_{s_j}$%\linebreak%$\wedge$ $h(\{v[s_j][f_i]\,|\,\forall f_i\}) = h(\vec{dv})$
	$step[s_j] :=$ rp-prepared
	%\pseudocode{Broadcast }%$\langle \msgcommit{}, v_{s_j}, s_j, r_i, h(\vec{dv}) \rangle_{\sigma_{r_i}}$%\smallskip%
%\pseudocode{\textbf{Replica} }%$r_i$%\pseudocode{ \textbf{receives} }%$\langle \msgcommit{}, v_{s_j}, s_j, *, h(\vec{dv}) \rangle $%\linebreak\pseudocode{ \textbf{with identical} $h(\vec{dv})$ \textbf{from} }%$2f+1$%\pseudocode{ \textbf{replicas}}\label{code:agr-2f1-commits}%:
	%\lstbtt{pre:}% $step[s_j] =$ rp-prepared $\wedge$ $view[s_j] = v_{s_j}$%\linebreak%$\wedge$ $h(\{v[s_j][f_i]\,|\,\forall f_i\}) = h(\vec{dv})$
	$step[s_j] :=$ rp-committed
	%\pseudocode{Stop commit timeout}% $\Delta_{commit}$ %\pseudocode{for slot}% $s_j$
	$exec[s_j] := \langle{}pr[s_j], \cup D_{f_i} \in \vec{dv} \rangle$
	%\pseudocode{Forward }%$\langle pr[s_j], D, s_j \rangle$%\pseudocode{ to execution, with }%$D :=\,\cup D_{f_i} \in \vec{dv}$%\label{code:agr-rp-exec}\vspace{2mm}\hrule\begin{center}\textbf{View Change}\end{center}\hrule\vspace{2mm}%
%\pseudocode{\textbf{Move to new view} }%$v_{s_j}$%\pseudocode{ \textbf{for slot} }%$s_j$%\pseudocode{ \textbf{at replica} }%$r_i$:
	%\pseudocode{If propose timeout $\Delta_{propose}$ for slot $s_j$ is active, trigger its expiry}\label{code:agr-propose-on-vc}%
	%\pseudocode{Stop commit/vc timeout}% $\Delta_{commit} $ and $\Delta_{vc}$ %\pseudocode{for slot}% $s_j$
	$dp := \langle p[s_j], pr[s_j] \rangle$; $\vec{dv} := \{v[s_j][f_i]\,|\,\forall f_i\}$
	/* Update certificate if current view fp-verified / rp-prepared%\,%*/
	%\pseudocode{If }%$step[s_j] \in$ {fp-verified, fp-committed}:%\label{code:agr-vc-cert-start}%
		$cert[s_j] := \langle \textsc{FPC}, dp, \vec{dv}, -1 \rangle$
	%\pseudocode{Else If }%$step[s_j] \in$ {rp-prepared, rp-committed}:
		$\vec{prep} :=$ %\pseudocode{set of}% $2f+1$ %\msgprepare{}s \pseudocode{with }%$h(\vec{dv})$
		$cert[s_j] := \langle \textsc{RPC}, dp, \vec{dv}, \vec{prep}, view[s_j] \rangle$%\label{code:agr-vc-cert-end}%
	$view[s_j] := v_{s_j}$
	$views[s_j][r_i] := v_{s_j}$
	$step[s_j] :=$ view-change
	%\pseudocode{Start query execute timeout}% $\Delta_{query-exec}$ %\pseudocode{for slot}% $s_j$
	%\pseudocode{Broadcast }%$\langle \msgvc, v_{s_j}, s_j,  r_i, cert[s_j]\rangle _{\sigma_{r_i}}$%\smallskip%
%\pseudocode{\textbf{Replica }%$r_i$%\pseudocode{ \textbf{receives} }%$\langle \msgvc, v_{s_j}, s_j,  r_k, * \rangle$:
	%\lstbtt{pre:}% $v_{s_j} > views[s_j][r_k]$%\hfill%/* View of a replica must only increase%\,%*/
	$views[s_j][r_k] := v_{s_j}$
	/* Move to f+1-highest known  view %\,%*/
	$vn := f+1$%\pseudocode{-highest in }%$\{views[s_j][r_l]\,|\,\forall r_l\}$%\label{code:agr-vc-inc-start}%
	%\pseudocode{If }%$vn > view[r_i]$:
		%\pseudocode{Move to new view }%$vn$%\hfill%/* Sends new %\msgvc%message%\,%%\label{code:agr-vc-inc-end}%*/%\smallskip%
$co := (s_j.co\,+\,\,$max$(0, v_{s_j}))$ mod $N$%\linebreak%
%\pseudocode{\textbf{View-change coordinator} }%$co$%\pseudocode{ \textbf{for view} }%$v_{s_j}$%\pseudocode{ \textbf{receives valid} }%$VCS:=\{\langle \msgvc, v_{s_j}, s_j, *, *\rangle\}$%\pseudocode{ \textbf{from} }%$2f+1$%\pseudocode{ \textbf{replicas}}%:
	%\lstbtt{pre:}% $step[s_j] =$ view-change $\wedge$ $view[s_j] = v_{s_j}$
	assert $\forall VC \in VCS: VC$ %\pseudocode{is valid}\linebreak%$\wedge\, (VC.cert = \emptyset \vee VC.cert.view \leq VC.v_{s_j})$
	%\pseudocode{Pick $dp$, $\vec{dv}$ from}%%\label{code:agr-pick-cert-start}%
			%\pseudocode{reconciliation-path result for highest view  if RPC certificate exists}%
			%\pseudocode{fast-path result\hspace{32.1mm}if FPC certificate exists}%
			%\pseudocode{}%null%\hspace{43mm}\pseudocode{otherwise}%%\label{code:agr-pick-cert-end}%
	%\pseudocode{Broadcast }%$\langle \msgnewview{}, v_{s_j}, s_j, co, dp, \vec{dv}, VCS\rangle _{\sigma_{co}}$%\smallskip%
%\pseudocode{\textbf{Replica }%$r_i$%\pseudocode{ \textbf{receives valid} }%$VCS:=\{\langle \msgvc, v_{s_j}, s_j, *, *\rangle\}$%\pseudocode{ \textbf{from} }%$2f+1$%\pseudocode{ \textbf{replicas}}%:
	%\lstbtt{pre:}% $step[s_j] =$ view-change $\wedge$ $view[s_j] = v_{s_j}$
	%\pseudocode{Start VC timeout}% $\Delta_{vc}$ %\pseudocode{for slot}% $s_j$
	%\pseudocode{Stop query execute timeout}% $\Delta_{query-exec}$ %\pseudocode{for slot}% $s_j$
%\textbf{Timeout }%$\Delta_{vc}$%\textbf{ for slot }%$s_j$%\textbf{ expires:}%
	%\pseudocode{Move to new view }%$v_{s_j}+1$
%\pseudocode{\textbf{Replica }%$r_i$%\pseudocode{ \textbf{receives} }%$\langle \msgnewview, v_{s_j}, s_j,  co, dp, \vec{dv}, VCS \rangle$:
	%\lstbtt{pre:}% $step[s_j] =$ view-change $\wedge$ $view[s_j] = v_{s_j}$
	assert $co$%\pseudocode{ is view-change coordinator for view }%$v_{s_j}$
	assert $\forall VC \in VCS: VC$ %\pseudocode{is valid}%
	assert $dp, \vec{dv}$%\pseudocode{ correctly picked based on }%$VCS$
	$\langle p[s_j], pr[s_j]\rangle := dp$%\label{code:agr-psj-vc}%
	$v[s_j][*] := \emptyset$%\hfill%/* Cleanup $\msgverify{}s$%\,%*/
	$\forall dv \in \vec{dv}: v[s_j][dv.f_i] := dv$
	%\pseudocode{If }%$s_j.i = r_i \wedge dp = $ null%\pseudocode{:}%
		permute-fast-quorum()
		%\pseudocode{Re-propose request in a new slot}%%\label{code:agr-repropose-after-null}%
	%\pseudocode{Start commit timeout}% $\Delta_{commit}$ %\pseudocode{with reduced timeout $\Delta_{vc-commit}$}% %\pseudocode{for slot}% $s_j$
	%\pseudocode{Enter reconciliation path}%
%\textbf{Timeout }%$\Delta_{query-exec}$%\textbf{ for slot }%$s_j$%\textbf{ expires:}%%\label{code:agr-exec-start}%
	%\pseudocode{Broadcast }%$\langle \msgqueryexec, s_j, r_j\rangle_{\sigma_{r_i}}$%\pseudocode{ to all replicas}%%\label{code:agr-broadcast-query-exec}%
%\pseudocode{\textbf{Replica }%$r_i$%\pseudocode{ \textbf{receives} }%$\langle \msgqueryexec, s_j, r_j\rangle$:
	%\lstbtt{pre:}% $exec[s_j] \neq \emptyset$
	$\langle dp, D \rangle := exec[s_j]$
	%\pseudocode{Send }%$\langle \msgexec,s_j,r_i, dp, D\rangle_{\sigma_{r_i}}$%\pseudocode{ to replica $r_j$}%
%\pseudocode{\textbf{Replica }%$r_i$%\pseudocode{ \textbf{receives} }%$\langle \msgexec,s_j,*, dp, D\rangle$ %\pseudocode{ \textbf{from $f+1$ replicas:}}%
	%\lstbtt{pre:}% $exec[s_j] = \emptyset$
	$exec[s_j] := \langle dp, D\rangle$
	%\pseudocode{Forward $\langle dp, D, s_j \rangle$ to execution with dependencies $D$}\label{code:agr-exec-exec}%%\label{code:agr-exec-end}%
\end{lstlisting}
\hrule\vspace{1mm}

\newtheorem{theorem}{Theorem}[section]
\newtheorem{conjecture}[theorem]{Conjecture}
\newtheorem{proposition}[theorem]{Proposition}
\newtheorem{lemma}[theorem]{Lemma}
\newtheorem{corollary}[theorem]{Corollary}
\newtheorem{example}[theorem]{Example}
\newtheorem{definition}[theorem]{Definition}
\newtheorem{remark}[theorem]{Remark}

\subsection{Proof}
\label{sec:proof}
We show that \system provides the following properties:
\begin{itemize}
	\item \textbf{Validity}: Only correctly signed client requests are executed.
	\item \textbf{Consistency}: Two correct replicas commit the same request and dependencies for a slot. \cite{moraru13there}
	\item \textbf{Execution Consistency}: Two interfering requests are executed in the same order on all correct replicas. \cite{moraru13there}
	\item \textbf{Linearizability}: If two interfering requests are proposed one after another, such that the first request is executed at some correct replica before the second request is proposed, then all replicas will execute the requests in that order.
	\item \textbf{Agreement Liveness}: In synchronous phases a client request will eventually commit. \cite{moraru13there}
	\item \textbf{Execution Liveness}: In synchronous phases a client will eventually receive a result.
\end{itemize}

We write $p^i$ to refer to a variable~$p$ from the perspective of replica~$r_i$.

We make the following standard cryptography assumptions:
A message $m$ signed by replica~$r_i$ is denoted as $\langle m \rangle_{r_i}$.
All replicas are able to verify each other's signatures.
A malicious replica is unable to forge signatures of correct replicas.
All replicas drop messages without a valid signature.

By $h(m)$ we refer to the hash of a message $m$ calculated using a collision resistant hash function, that is $h(m) \neq h(m') \Rightarrow m \neq m'$.

Messages for a slot are delivered eventually by retransmitting them unless the slot was garbage collected in the meantime.
That is we assume reliable point-to-point connections between all replicas until slots are garbage collected.
Once a replica has successfully completed a view change for a slot, then it is no longer necessary to retransmit messages for earlier views. It is also not necessary to retransmit \msgpropose and \msgverify messages once a new view was entered for the slot. In addition, for each message type only the message from the highest view in which the message type was sent has to be retransmitted.

We first show the properties for \system without checkpointing and extend the pseudocode and proofs to include checkpointing later on.

\subsubsection{Validity}
\begin{theorem}[Validity]
	Only correctly signed client requests are executed.
\end{theorem}

\begin{IEEEproof}
The execution algorithm only executes committed requests.
For a client request to execute it must reach Line~\ref{code:agr-fp-exec} or \ref{code:agr-rp-exec} in the Agreement Protocol in variable~$p[s_j]$ or be received via \msgexec messages in Line~\ref{code:agr-exec-exec}.
$p[s_j]$ is set in
\begin{itemize}
	\item Line~\ref{code:agr-psj-coord} and \ref{code:agr-psj-follower}: The validity of the client request was verified after receiving.
	\item Line \ref{code:agr-psj-vc}: The value can be \texttt{null} or a value from a certificate.
	As each valid certificate contains $2f$~\msgverify{}s from the initial view, one must be from a correct replica and as a correct replica only creates a valid \msgverify once after verifying the client request (L.~\ref{code:agr-send-verify}), the request must be correct.
	Otherwise, a correct replica must have created two \msgverify{}s which yields a contradiction.
\end{itemize}

Requests received via \msgexec are only forwarded to the execution if a replica receives $f+1$~matching \msgexec{}s.
Thus, at least one \msgexec is from a correct replica which must have processed the request according to one of the two previous cases.

The \texttt{null} request is skipped during execution and thus only correctly signed client requests are executed.
\end{IEEEproof}

\subsubsection{Consistency}
We first establish some notation:
\begin{definition}
	A slot~$s_j$ is \emph{verified} if a correct replica collects a valid \msgpropose~$dp$, $2f$~valid \msgverify{}s from different replicas with matching~$h(dp)$ and each \msgverify is from a replica in the fast-path quorum~$dp.F$.
\end{definition}
\begin{definition}
	A slot~$s_j$ is \emph{fp-verified} if a correct replica \emph{verified} it and each dependency in the \msgverify{}s occurs at least $f+1$ times.
\end{definition}
\begin{definition}
	A slot~$s_j$ is \emph{fp-committed} if a correct replica collects $2f+1$~\msgfpcommit{}s from different replicas with matching~$h(\vec{dv})$.
\end{definition}
\begin{remark}
	Note that \emph{fp-committed} implies \emph{fp-verified} as \msgfpcommit{}s are only sent by replicas which \emph{fp-verified} the slot.
\end{remark}
\begin{definition}
	A slot~$s_j$ is \emph{rp-verified} if a correct replica \emph{verified} it and it is not \emph{fp-verified}.
\end{definition}
\begin{definition}
	A slot~$s_j$ is \emph{rp-prepared} if a correct replica collects $2f+1$~\msgprepare{}s from different replicas with matching~$h(\vec{dv})$.
\end{definition}
\begin{definition}
	A slot~$s_j$ is \emph{rp-committed} if a correct replica collects $2f+1$~\msgcommit{}s from different replicas with matching~$h(\vec{dv})$.
\end{definition}
\begin{remark}
	Similar to \emph{fp-committed}, \emph{rp-committed} implies \emph{rp-prepared} and \emph{rp-prepared} implies \emph{rp-verified}.
\end{remark}
\begin{definition}
	A slot~$s_j$ is \emph{committed} if a correct replica \emph{fp-committed} or \emph{rp-committed} it.
\end{definition}
\begin{theorem}[Consistency]
	\label{thm:consistency}
	Two correct replicas commit the same request and dependencies for a slot.
\end{theorem}
We first proof the following auxiliary lemmas.

\begin{lemma}
	\label{lemma:inv-fp-rp}
	A slot~$s_j$ cannot both be \emph{fp-committed} and \emph{rp-prepared} in $view=-1$.
\end{lemma}
\begin{IEEEproof}
	By contradiction.
	Assume that a slot is both \emph{fp-committed} and \emph{rp-prepared} in $view=-1$.
	As the slot was \emph{rp-prepared} a correct replica received $2f+1$~\msgprepare{}s~(L.~\ref{code:agr-2f1-prepares}).
	This requires $f+1$~correct replicas to have entered the reconciliation path (via L.~\ref{code:agr-jump-rp} and \ref{code:agr-enter-rp}).
	To be \emph{fp-committed}, another replica must have received $2f+1$~\msgfpcommit{}s.
	As sending a \msgfpcommit~(L.~\ref{code:agr-send-fpcommit}) and entering the reconciliation path~(L.~\ref{code:agr-jump-rp}) are mutually exclusive, a correct replica must have done both which yields a contradiction.
\end{IEEEproof}
\begin{lemma}
	\label{lemma:secure-rpc}
	The content of a reconciliation-path certificate (RPC) cannot be manipulated.
\end{lemma}
\begin{IEEEproof}
	As each protocol phase includes hashes of the previous phase, faulty replicas can only manipulate the last round of messages included in a certificate.
	For an RPC only the \msgprepare{}s can be manipulated, however, these only confirm the values selected by the \msgpropose and \msgverify{}s.
\end{IEEEproof}
\begin{lemma}
	A faulty replica can only create a manipulated but valid fast-path certificate (FPC) if not \emph{fp-committed}.
\end{lemma}
\begin{IEEEproof}
	A faulty replica could construct a faulty FPC using manipulated \msgverify{}s, allowing the replica to include manipulated dependency sets.
	A replica only takes the fast path, if each dependency was reported in at least $f+1$~\msgverify{}s~(L.~\ref{code:agr-f1-deps}).
	This requirement is also necessary for an FPC to be valid.
	
	Only a single \msgpropose can be \emph{verified}, which requires $2f$ matching \msgverify{}s, as each correct replica only sends a \msgverify for the first \msgpropose for a slot.
	Thus correct replicas use the same fast-path quorum~$F$ to create an FPC and all valid FPCs must use the same $F$.
	To change the dependency sets faulty replicas only have the option to create manipulated \msgverify{}s.
	
	We now proof the Lemma by contradiction.
	Assume \emph{fp-committed} holds.
	Then a correct replica has received $2f$~\msgverify{}s in which each dependency is part of $0$ (nonexistent dependency) or $f+1$~\msgverify{}s.
	\begin{itemize}
		\item $0$~occurrences: A manipulated FPC can either not include the dependency, in which case the FPC is unchanged.
		Or include a new dependency up to $f$ times, which causes the FPC to become invalid.
		\item $f+1$~or more occurrences:
		A manipulated FPC can either include the existing dependency at least $f+1$ times, in which case the outcome of applying the FPC is unchanged.
		Or include a dependency only between $1$ and $f$ times, which causes the FPC to become invalid.
	\end{itemize}
\end{IEEEproof}
\begin{lemma}
	\label{lemma:secure-fpc}
	A manipulated FPC can only be used if neither \emph{fp-committed} nor \emph{rp-committed}.
\end{lemma}
\begin{IEEEproof}
	If \emph{fp-committed}, then no manipulated FPC can exist.
	If \emph{rp-committed}, at least $f+1$~correct replicas have \emph{rp-prepared} and thus at least one RPC is contained in one of the $2f+1$~\msgvc{}s required for the view change.
	Thus the FPC is ignored.
	As \emph{fp-committed} and \emph{rp-prepared} are mutually exclusive and \emph{rp-committed} implies \emph{rp-prepared}, no valid FPC can exist.
\end{IEEEproof}

Now we proof Theorem~\ref{thm:consistency}:
\begin{IEEEproof}
\textbf{Case 1:} A replica~$r_i$ commits $\langle c, D, s_j\rangle$ via the fast-path~(L.~\ref{code:agr-fp-exec}). $c$ is the request committed with dependencies~$D$ for slot~$s_j$.
\begin{itemize}
	\item Case 1.1: Another replica~$r_k$ commits $\langle c', D', s_j\rangle$ with $c\neq c' \vee D \neq D'$ via the fast-path.
	\begin{itemize}
		\item Case~$c \neq c'$:
			\begin{IEEEproof}
			Then $p^i[s_j] \neq p^k[s_j]$, as $c\neq c'$.
			$h(p[s_j])$ is part of the \msgverify{}s.
			 Therefore, $h(\vec{dv})$ must differ.
			 Then $r_i$ and $r_k$ each need $2f+1$~\msgfpcommit{}s with different $h(\vec{dv})$, which due to the properties of a Byzantine majority quorum would require a correct replica to send two \msgfpcommit{}s, which yields a contradiction.
			 \end{IEEEproof}
		\item Case $D \neq D'$:
			\begin{IEEEproof}
			With $D := \cup D_{f_i} \in \vec{dv}$ it follows, that for a differing fast-path quorum~$F$ or dependency sets~$D_{f_i}$, $r_i$ and $r_k$ must use different $h(\vec{dv})$.
			Now, the proof of the previous case applies.
			\end{IEEEproof}
	\end{itemize}
	\item Case 1.2: $r_k$ commits in view~$-1$ via the reconciliation path.
		\begin{IEEEproof}
		Then \emph{rp-committed} holds which requires $2f+1$~\msgcommit{}s (L.~\ref{code:agr-2f1-commits}) of which $f+1$ must originate from correct replicas.
		This implies \emph{rp-prepared} which according to Lemma~\ref{lemma:inv-fp-rp} conflicts with \emph{fp-committed}.
		\end{IEEEproof}
	\item Case 1.3: $r_k$ commits $\langle c', D', s_j\rangle$ with $c\neq c' \vee D \neq D'$ in $view \geq 0$ via the reconciliation path.\\
		Deferred to Case 3.
\end{itemize}
\textbf{Case 2:} A replica~$r_i$ commits $\langle c, D, s_j\rangle$ via the reconciliation path in $view=-1$~(L.~\ref{code:agr-rp-exec}).
\begin{itemize}
	\item Case 2.1: $r_k$ commits $\langle c', D', s_j\rangle$ with $c\neq c' \vee D \neq D'$ via the fast-path.
		\begin{IEEEproof}
		See Case 1.2.
		\end{IEEEproof}
	\item Case 2.2: $r_k$ commits in view~$-1$ via the reconciliation path.
		\begin{IEEEproof}
		This requires two sets of $2f+1$~\msgprepare{}s with different $h(\vec{dv})$ which would require a correct replica to send  two different \msgprepare{}s.
		\end{IEEEproof}
	\item Case 2.3: $r_k$ commits $\langle c', D', s_j\rangle$ with $c\neq c' \vee D \neq D'$ in $view \geq 0$ via the reconciliation path.\\
		Deferred to Case 3.
\end{itemize}
\textbf{Case 3:} A replica~$r_k$ commits $m' := \langle c', D', s_j\rangle$ via the reconciliation path in $view\geq 0$.
\begin{IEEEproof}
	We proof this by induction: Once a replica commits~$m := \langle c, D, s_j\rangle$, with $c\neq c' \vee D \neq D'$, in some $view$, then no replica can commit or prepare a different result~$m'$ in views~$> view$.
	
	Base case: $view' = view + 1$:
	
	Assume that $m$ committed in $view$ and that $m'$ prepares / commits in $view'$.
	A correct replica only decides a result in view~$view'$ after receiving a valid \msgnewview.
	No manipulated RPC and FPC are used according to Lemma~\ref{lemma:secure-rpc} and \ref{lemma:secure-fpc}.
	\begin{itemize}
		\item Case $view=-1\,\wedge$ \emph{fp-committed}:
			No RPC can exist, as \emph{fp-committed} and \emph{rp-prepared} are mutually exclusive.
			As the fast-path committed, at least $f+1$~correct replicas have \emph{fp-verified} the slot.
			These will include an FPC in their \msgvc.
			As the view-change coordinator has to wait for $2f+1$~\msgvc{}s, at least one \msgvc{} will include the FPC.
			The FPC would include $m$, which contradicts the assumption.
		\item Case $view=-1\,\wedge$ \emph{rp-committed} :
			$f+1$~correct replicas must be \emph{rp-prepared} and thus provide the view-change coordinator with an RPC, which must be included in the \msgnewview.
			No correct replica can be \emph{fp-committed}, as it is mutually exclusive with \emph{rp-prepared}.
			Therefore, if valid RPC and FPC exist, then the FPC is from a faulty replica and must be ignored.
			Thus, the selected RPC contains $m$ which yields a contradiction.
		\item Case $view \geq 0$:
			The slot must have \emph{rp-committed} and thus, similar to the previous case, the view change correctly selects the RPC.
			Therefore, the RPC contains $m$ which yields a contradiction.
	\end{itemize}
	
	Induction step: $view' > view + 1$:
	
	To commit a slot, $2f+1$ replicas have to send a \msgfpcommit or \msgcommit .
	One \msgvc with a corresponding certificate from a correct replica must be part of the $2f+1$~\msgvc messages.
	A correct replica always sends its newest certificate~(L.~\ref{code:agr-vc-cert-start}-\ref{code:agr-vc-cert-end}), and therefore one of the \msgvc{}s used by the view-change coordinator includes a certificate from the highest view $v_{max}$ in which a request has committed.
	\begin{itemize}
		\item Case $v_{max} \geq 0$:
		Thus the reconciliation path must have committed in $v_{max}$, and therefore the correct certificate is selected~(L.~\ref{code:agr-pick-cert-start}-\ref{code:agr-pick-cert-end}).
		\item Case $v_{max} = -1$:
		The existence of an RPC shows that not \emph{fp-committed} and thus the RPC must be selected.
		If only an FPC exists, then not \emph{rp-committed} and therefore it is valid to select the FPC.
	\end{itemize}
\end{IEEEproof}
\textbf{Case 4:} A replica~$r_k$ commits $\langle dp, D, s_j\rangle$ after receiving $f+1$~valid $\langle \msgexec{}, s_j, *, dp, D\rangle$ messages~(Line~\ref{code:agr-exec-start}-\ref{code:agr-exec-end}).
This case allows lagging replicas to catch up and learn the agreement result as described in Section~\ref{sec:progress-guarantees}.
\begin{IEEEproof}
	At least one of the \msgexec~messages is from a correct replica, which either has committed the slot itself in which case the previous cases apply.
	Or the correct replica has learned from another correct replica that the slot was committed.
\end{IEEEproof}

The cases are exhaustive.
\end{IEEEproof}

\subsubsection{Execution Consistency}
\label{sec:appendix-execution consistency}
Execution pseudo code
\vspace{1mm}\hrule\vspace{-1mm}
\begin{lstlisting}[name=pseudocode]
%\pseudocode{\textbf{Variables at each replica}}%:
$k$%\hfill% /* Size of execution window%\,%*/
$committed$, $executed$ /* Sets containing all slots which have been commmitted or executed so far%\,%*/
$exp(r_i) := \min \{v_{min}\,|\, v_{min} \notin executed \vee v_{min}.i = r_i\}$%\hfill%/* First not executed request for replica $r_i$, defines the lower bound of the execution window%\,%*/
$exp_k := \{v | v.s_i < exp(v.i)+ k\}$%\hfill%/* Executed slots and slots in execution window %\,%*/
$rhist[*]:=\bot$%\hfill%/* History variable for dependency graph calculation%\,%*/%\vspace{2mm}\hrule\begin{center}\textbf{Request Execution}\end{center}\hrule\vspace{2mm}%
/* Calculate dependency graph for slot $v$%\,%*/
$rdeps(v)$:
	$ D' := \{v\}$
	%\pseudocode{While }%$D \neq D'$:
		$D := D'$
		%\pseudocode{For }%$v \in D$:
			%\pseudocode{If }%$v \notin executed$:
				$D' := D' \bigcup_{d\in deps(v)} (v \rightarrow d) \cup \{d\}$%\label{code:exec-rdeps-deps}%
			%\pseudocode{Else}%:
				$D' := D' \cup rhist[v]$
	%\pseudocode{Return }%$D$
	
/* Calculate dependency graph for slot $v$. Excludes slots outside the execution window%\,%*/
$rdeps_{exp}(v)$:
	$D' := \{v\} \cap exp_k$
	%\pseudocode{While }%$D \neq D'$:
		$D := D'$
		%\pseudocode{For }%$v \in D$:
			%\pseudocode{If  }%$v \notin executed$:
				$D' := D' \bigcup_{d\in deps(v), d \in exp_k} (v \rightarrow d) \cup \{d\}$%\label{code:exec-rdeps-exp-deps}%
			%\pseudocode{Else}%:
				$D' := D' \cup rhist[v]$
	%\pseudocode{Return }%$D$%\linebreak\linebreak\linebreak%
%\pseudocode{While True:}%
	%\pseudocode{Update slots committed in the meantime}%
	/* repeat loop until no further suitable $v$ exists%\,%*/
	%\pseudocode{For all}\,%$v \in (exp_k \setminus executed) \wedge rdeps(v) \subseteq (committed \cap exp_k)$:%\label{code:exec-regular-check}%
		$\vec{sc} := $%\pseudocode{ find not executed strongly-connected components in $rdeps(v)$ in inverse topological order}%
		%\pseudocode{For}% $sc \in \vec{sc}$:
			/* normal case execution%\,%*/%\label{code:exec-normal-execution}%
			execute($sc$, $rdeps(sc)$)
	%\pseudocode{For all}% $v\in (exp_k \setminus executed) \wedge rdeps_{exp}(v) \subseteq committed$:%\label{code:exec-special-check}%
		$\vec{sc} := $%\pseudocode{ find not executed strongly-connected components in $rdeps_{exp}(v)$ in inverse topological order}%
		/* unblock execution case%\,%*/%\label{code:exec-unblock-execution}%
		execute($\vec{sc}[0]$, $rdeps_{exp}(\vec{sc}[0])$)

%\textbf{\lstbtt{execute($\vec{v}$, $d$):}}%
	%\pseudocode{For}% $c \in sort(\vec{v})$:
		%\pseudocode{Execute request}% $c$
		$rhist[v] := d$

%\textbf{\lstbtt{sort($\vec{v}$):}}%
	%\pseudocode{Return $\vec{v}$ sorted by sequence numbers $v.seq$ and use replica id $v.i$ as tie breaker}%
\end{lstlisting}
\vspace{1mm}\hrule\vspace{1mm}

Similar to the Execution Consistency property in EPaxos~\cite{moraru13there}, we show:
\begin{theorem}[Execution Consistency]
	\label{thm:sscc-trace-completeness}
	All replicas execute all pairs of committed, conflicting requests in the same order.
\end{theorem}

The relation $conflict(a, b)$ states that two requests $a$ and $b$ conflict with each other.
The dependencies which were committed for a slot~$a$ are given by $deps(a)$.
We first show that conflicting requests have dependencies on each other, before showing that conflicting requests are executed in the same order on all replicas.

\begin{lemma}
	If $conflict(a, b)$ then, $a$ has a dependency on $b$ in $deps(a)$ or the other way around. 
\end{lemma}
\begin{IEEEproof}
	For a request~$r$ one \msgpropose and $2f$~\msgverify, that is in total $2f+1$~replicas, provide dependencies for the request.
	
	For $a$ and $b$ at least one correct replica~$r_i$ receives both requests.
	\begin{itemize}
		\item $r_i$ receives $a$ before $b$: Then $a\in deps(b)$.
		\item $r_i$ receives $b$ before $a$: Then $b\in deps(a)$.
	\end{itemize}

	Therefore, the dependency is included when the slot is \emph{committed} via the fast or reconciliation path in $view=-1$.
	When a replica \emph{rp-prepares} the slot, then its RPC must by construction also include the dependency.
	As an FPC must include $f$~\msgverify{}s and a \msgpropose{} from correct replicas or $f+1$~\msgverify{}s from correct replicas, one of these messages includes the dependency.
	This is the case as a faulty replica cannot change the fast-path quorum~$F$ afterwards and thus cannot change which replicas contribute to an FPC.
	
	In case no FPC or RPC is part of the view change, then a \texttt{null} request is selected.
	As that request does not conflict with any other request, no dependencies are required.
	
	Note that the requirement for an FPC or RPC which include \msgpropose and \msgverify{}s ensures that only the request coordinator can propose a request for the slot.
\end{IEEEproof}

Next, we show that conflicting requests are executed in the same order on all replicas.
We start with several definitions used in the following:
\begin{definition}[SCC trace]
A strongly-connected components~(SCC) trace~$t$ is a 0-based vector consisting of executed SCCs in the order of their execution.
We write $t^i$ to refer to the trace belonging to a replica~$r_i$.
\end{definition}
\begin{definition}[SSCC trace]
A special-case strongly-connected component~(SSCC) trace~$\hat{t}$ is the subset of an SCC trace~$t$ containing only the SSCCs, that is all SCCs which were executed via the unblock execution case (Line~\ref{code:exec-unblock-execution}).
\end{definition}
\begin{definition}[Regular SCC]
A regular SCC is one which was executed via the normal case execution (Line~\ref{code:exec-normal-execution}).
\end{definition}
All slots executed as part of a trace are given by $flatten(t) = \{v\,|\,v \in s, s \in t\}$.

For a slot~$v$ we write $v.i$ to refer to replica~$i$ which is the request coordinator for that slot.
And $v.seq$ to access the attached counter / sequence number.
For two slots~$v_1$ and $v_2$ with $v_1.i = v_2.i$, we use $v_1 < v_2$ as shorthand for $v_1.seq < v_2.seq$.

We write $a \rightarrow b$ if $a$ directly depends on $b$, that is $b \in deps(a)$.
(Logical implications are written as $P \Rightarrow Q$.)
$a \leadsto b$ also includes transitive dependencies, that is $a \leadsto b \Leftrightarrow a \rightarrow b \vee  a \rightarrow v_1 \rightarrow \ldots \rightarrow v_n \rightarrow b$, with $n\in \mathds{N}$.

$rdeps(v)$ and $rdeps_{exp}(v)$ define dependency graphs starting from a slot $v$.
They return a graph consisting of slots and edges $v_1 \rightarrow v_2$ between slots in these graphs.
By construction all slots and edges in the graph are reachable from $v$.

\begin{corollary}
\label{cor:scc-order}
An SCC trace fully defines the order in which requests are executed.
\end{corollary}
\begin{IEEEproof}
Slots can only be executed via $execute(\vec{v},d)$, which groups requests by SCCs. As the execution algorithm filters out executed slots, each slot is only executed once and can thus only be part of one SCC.
Note that the inverse topological sorting ensures that slots in an SCC can only depend on the SCC itself or earlier SCCs.
Requests within an SCC are sorted before execution, which yields a stable order.
\end{IEEEproof}
Note that by definition $executed^i = flatten(t^i)$.

We first show the following supporting Lemma:
\begin{lemma}
	\label{lemma:all-equal-rhist}
	Assume correct replicas~$r_i$ and $r_j$ have two traces~$t^i$ and $t^j$ where $\hat{t^i}=\hat{t^j}$. Then $\forall v\in flatten(t^i) \cap flatten(t^j): rhist^i[v] = rhist^j[v]$, where $v$ is a slot in an SCC.
\end{lemma}
That is for all slots executed at both replica~$r_i$ and $r_j$ ($\forall v\in flatten(t^i) \cap flatten(t^j)$), the dependency sets used for ordering are identical on these replicas for each of these slots.

We proof Lemma~\ref{lemma:all-equal-rhist} in multiple steps by induction.
Base case of Lemma~\ref{lemma:all-equal-rhist}: The Lemma applies for an SSCC trace~$\hat{t}$ with length 0, that is trace~$t$ only contains regular SCCs.

\begin{lemma}
	\label{lemma:scc-equal-rdeps}
	When $\forall v\in flatten(t^i) \cap flatten(t^j): rhist^i[v] = rhist^j[v]$ before executing an SCC~$s$ via the normal case then $\forall v' \in s: rdeps^i(v') = rdeps^j(v')$.
\end{lemma}
\begin{IEEEproof}
	By construction an SCC is only executed after all its slots were committed, therefore $deps^i(v) = deps^j(v) = deps(v)$ are identical on all replicas for any slot~$v$ used during the calculation of $rdeps^i(v')$ and $rdeps^j(v')$.
	
	We set $v_1$ to be an arbitrary slot of SCC~$s$.
	As by assumption no slot of the SCC was executed before, $rdeps(v_1)$ must expand all slots~$v_s$ of the SCC~$s$ via $deps(v_s)$~(Lines~\ref{code:exec-rdeps-deps} and \ref{code:exec-rdeps-exp-deps}).
	As all dependencies of an SCC are executed before the SCC, then for slot~$v_d$ in one of these dependencies, $rhist[v_d]$ is identical on all replicas.
	Therefore $rdeps^i(v_1) = rdeps^j(v_1)$.

	This completes the proof for SCCs of size 1.
	In the following we consider SCCs consisting of at least two slots and show that if two SCCs at different replicas have at least one slot in common, then the SCCs are identical.
	By definition $\forall v_2 \in SCC, v_1\neq v_2: v_1 \leadsto v_2 \wedge v_2 \leadsto v_1$.
	Thus $rdeps^i(v_1) = rdeps^i(v_2)$.
	
	Now assume that $s^i \neq s^j \wedge s^i \cap s^j \neq \emptyset$:
	W.l.o.g $v_1 \in s^i, \notin s^j$ and $v_2 \in s^i \cap s^j$.
	Then $v_1 \in rdeps^i(v_1) = rdeps^i(v_2) = rdeps^j(v_2)$ and therefore $v_1 \in s^j$ which yields a contradiction.
	Thus $s^i = s^j$.
\end{IEEEproof}
In the following we use $rdeps(v)$ and $rdeps(s)$ for slot $v\in \textrm{SCC~} s$ interchangeably.

\begin{lemma}
	\label{lemma:t0-equal-weak-rhist}
	Assume replica~$r_i$ and $r_j$ have two traces~$t^i$ and $t^j$ which only contain SCCs executed via the normal case,  that is $|\hat{t^i}|=0$ and $|\hat{t^j}|=0$. Then $\forall v\in flatten(t^i) \cap flatten(t^j): rhist^i[v] = rhist^j[v]$, where $v$ is a slot in an SCC.
\end{lemma}
\begin{IEEEproof}
	W.l.o.g. we assume $|t^i| = |t^j|$. A short trace can be padded with empty "SCCs" which are equivalent to a no-op.

	Base case $|t^i| = 0$:
	$flatten(t^i) \cap flatten(t^j) = \emptyset$.
	
	Induction step $|t'^i| = |t^i|+1$:
	We define $s^i := t'^i[|t'^i|-1]$ to be the last element in $t'^i$. We only discuss $s^i$, the same arguments apply for an $s^j$ with swapped $i$ and $j$.
	
	\textbf{Case 1}: $s^i \in set(t'^i) \cap set(t'^j)$: 
	Both $t'^i$ and $t'^j$ contain $s=s^i$.
	As the SCCs are sorted in inverse topological order, $s$ is only executed after all SCCs on which $s$ depends were executed first, that is $rdeps^i(s)$ can only contain slots~$v \in s \cup executed$.
	Note that all slots in the SCC~$s$ must already be committed (and not yet executed) as the SCC would not be executed otherwise.
	Per (Agreement) Consistency property the set $\mathcal{D} := \bigcup_{v\in s}deps(v)$ is identical on all replicas, and thus $s$ directly depends on the same slots on all replicas.
	These slots (and their SCCs) have been executed on both replicas~$r_i$ and $r_j$ and thus per induction assumption $\forall v \in \mathcal{D}: rhist^i[v] = rhist^j[v]$.
	Thus $rdeps^i(s) = rdeps^j(s)$ and therefore $rhist^i[s] = rhist^j[s]$.
	
	\textbf{Case 2}: $s^i \notin set(t'^i) \cap set(t'^j) \Leftrightarrow s^i \notin t'^j$:
	We show that for all $s^j \in t'^j$: if $s^i \neq s^j$,  then $s^i \cap s^j = \emptyset$.
	Assume that $s_j$ is the SCC with the lowest index in $t'^j$ with $s^i \cap s^j \neq \emptyset$.
	With Lemma~\ref{lemma:scc-equal-rdeps} this results in a contradiction.
	
	The cases are exhaustive.
	
	Thus, we have shown that $\forall v \in \{v | s \in set(t^i) \cap set(t^j), v \in s\}: rhist^i[v] = rhist^j[v]$.
	We also know that for $s_1\in t^i, s_2 \in t^j: s_1 \cap s_2 \neq \emptyset \Rightarrow s_1 = s_2$.
	To complete the proof of the Lemma we have to show that $LHS : = flatten(t^i) \cap flatten(t^j)$ equals $RHS := \{v | s \in set(t^i) \cap set(t^j), v \in s\}$.
	By construction $LHS \supseteq RHS$,  thus we have to show that $LHS \not\supset RHS$.
	
	Assume that is not the case: Pick $v$ such that $v \in LHS, v \notin RHS$. Then $v \in s_1 \in t^i, v \in s_2 \in t^j$.
	
	Case $s_1 = s_2$: Contradiction.
	
	Case $s_1 \neq s_2$: Thus $s_1 \cap s_2 = \emptyset$, which contradicts $v \in s_1, s_2$.
\end{IEEEproof}
\begin{corollary}
	\label{corollary:compact-deps}
	The compact dependency representation implicitly includes dependencies on all earlier slots of a replica. That is a dependency from slot~$a$ to slot $b$ ensures that $b \in deps(a) \Rightarrow deps(a) \supseteq deps(a, b) = \{ v | v.i = b.i \wedge v.seq \leq b.seq\}$.
\end{corollary}
\begin{lemma}
	\label{lemma:scc-sscc-uniqueness}
	A regular SCC is only executed by the normal case execution, whereas an SSCC is only executed by the unblock execution case.
\end{lemma}
\begin{IEEEproof}
	To reach a contradiction, assume that a regular SCC is executed via the unblock execution case.
	For this, we must find a slot~$v \in exp_k \cap committed, v \notin executed$ which satisfies the following condition:
	$rdeps_{exp}(v)\subseteq committed \wedge \neg (rdeps(v) \subseteq committed \cap exp_k)$.
	The first part of the condition ensures that the unblock execution case can execute (Line~\ref{code:exec-special-check}) and the second part ensures that the normal case execution does not apply~(Line~\ref{code:exec-regular-check}).
	$\Leftrightarrow rdeps_{exp}(v)\subseteq committed \wedge (rdeps(v)\not\subseteq committed \vee rdeps(v) \not\subseteq exp_k)$\\
	$\Leftrightarrow rdeps_{exp}(v)\subseteq committed \wedge (rdeps(v)\setminus  rdeps_{exp}(v)\not\subseteq committed \vee rdeps(v) \not\subseteq exp_k)$.
	
	We also make the following observation: $rdeps(v) \subseteq exp_k \Rightarrow rdeps(v) = rdeps_{exp}(v)$. If $rdeps(v) \subseteq exp_k$ then the check using $exp_k$ in $rdeps_{exp}$ never skips dependencies and therefore $rdeps(v) = rdeps_{exp}(v)$.

	Assume that the unblock case would execute while $rdeps(v) \subseteq exp_k$, then $rdeps(v)\setminus  rdeps_{exp}(v) = \emptyset$ which prevents the unblock case from executing.
	Thus, the unblock case only executes if $rdeps(v) \not\subseteq exp_k$.
	
	Due to the inverse topological sort order the SSCC $sc[0]$ in the unblock execution case must have $rdeps_{exp}(sc[0]) \setminus sc[0] \subseteq executed$ that is all dependencies except $sc[0]$ must be executed and $sc[0] \subseteq rdeps_{exp}(sc[0])$.
	As $sc[0] \subseteq exp_k$, $rdeps(sc[0]) \supset rdeps_{exp}(sc[0])$ and therefore $\exists x_a \in rdeps(sc[0]): (x_a \rightarrow x_e) \in rdeps(sc[0]) \wedge x_e \notin exp_k$. 
	Due to Corollary~\ref{corollary:compact-deps} $(x_a \rightarrow exp(x_e.i)) = (x_a \rightarrow x_r) \in rdeps_{exp}(sc[0]), x_r \in rdeps(sc[0])$.
	By definition~$x_e.seq - x_r.seq \geq k$ and therefore always $rdeps(sc[0]) \not\subset exp_k$.
	Thus, $sc[0]$ can never be executed via the normal case, which contradicts the assumption that $sc[0]$ is a regular SCC.
	
	For $v\in SSCC$, as $rdeps(v) \not\subset exp_k$ before executing $v$, it can never execute via the normal execution case.
\end{IEEEproof}
Lemma~\ref{lemma:scc-sscc-uniqueness}  allows strengthening Lemma~\ref{lemma:t0-equal-weak-rhist} to:
\begin{lemma}
	\label{lemma:t0-equal-rhist}
	Assume replica~$r_i$ and $r_j$ have two traces~$t^i$ and $t^j$ which only contain regular SCCs, that is $|\hat{t^i}|=0$ and $|\hat{t^j}|=0$. Then $\forall v\in flatten(t^i) \cap flatten(t^j): rhist^i[v] = rhist^j[v]$, where $v$ is a slot in a (regular) SCC.
\end{lemma}

\begin{lemma}
	\label{lemma:t0-equal-sccs}
	Assume replica~$r_i$ and $r_j$ have two traces~$t^i$ and $t^j$ which only contain regular SCCs, that is $|\hat{t^i}|=0$ and $|\hat{t^j}|=0$. If $s^i \neq s^j$, then $s^i \cap s^j = \emptyset$.
\end{lemma}
\begin{IEEEproof}
	Follows from the proof of Lemma~\ref{lemma:t0-equal-weak-rhist} and \ref{lemma:scc-sscc-uniqueness}.
\end{IEEEproof}

Induction step 1 of Lemma~\ref{lemma:all-equal-rhist}: The lemma applies for a fixed SSCC trace~$\hat{t'}$ with length $|\hat{t}|+1$ where the SSCC is the last element of trace~$t'$.
We refer to it as $\hat{s} := t'[|t'|-1] = \hat{t'}[|\hat{t'}|-1]$.

\begin{lemma}
	\label{lemma:t-hat-equal-rhist}
	$rhist^i[\hat{s}] = rhist^j[\hat{s}]$ for a fixed SSCC trace~$\hat{t' }$ with $\hat{s} := t'[|t'|-1]$.
\end{lemma}
\begin{IEEEproof}
For an SSCC to execute via the unblock case, the following must hold:
$rdeps_{exp}(\hat{s})\subseteq committed \wedge \neg (rdeps(\hat{s}) \subseteq committed \cap exp_k)$.
As shown in the proof of Lemma~\ref{lemma:scc-sscc-uniqueness} this requires $rdeps(\hat{s}) \not\subseteq exp_k$.

$rdeps_{exp}(\hat{s})$ depends on $deps(v)$, $rhist[v]$ and $exp_k$.
$deps(v)$  is identical across replicas by the (Agreement) Consistency property and $rhist[v]$ is identical across replicas as SCC dependencies are executed first and thus this follows from the induction assumption.
In the SSCC there must exist slots~$v_m \in exp_k$ with a dependency on a slot~$v_e \in deps(v_m), v_e \notin exp_k, \notin rdeps_{exp}(\hat{s})$.
We now show that $\forall v_m \in exp_k: \forall v_e \in deps(v_m), v_e \notin exp_k: exp^i(v_e.i) = exp^j(v_e.i)$.
Assume that this is not the case.

By Corollary~\ref{corollary:compact-deps}, $deps(v_m) \supseteq deps(v_m, v_e)$.
We set $v_r = min(\{d | d \in deps(v_m, v_e) \wedge d \notin executed\})$.
By definition $exp(v_e.i) = v_r$.
Note that $v_r$ must be part of SSCC as it would have to be $\in executed$ otherwise, which contradicts the definition of $v_r$.
That is for all replicas $r_l$ to which dependencies are removed in $rdeps_{exp}$, $exp(r_l)$ is defined by the slots in the SSCC.
Thus $rdeps_{exp}^i(\hat{s}) = rdeps_{exp}^j(\hat{s})$.

Due to Lemma~\ref{lemma:scc-sscc-uniqueness} the SSCC can only be executed via the unblock execution case.
\end{IEEEproof}
\begin{lemma}
	\label{lemma:t-hat-equal-slot-rhist}
	Assume replica~$r_i$ and $r_j$ have two traces~$t^i$ and $t^j$ where the SSCC~$\hat{s}$ is the last element and $\hat{t^i}=\hat{t^j}$. Then $ \forall v\in flatten(t^i) \cap flatten(t^j): rhist^i[v] = rhist^j[v]$, where $v$ is a slot in an SCC.
\end{lemma}
\begin{IEEEproof}
	By construction slots are only executed once, thus either $v \in flatten(\hat{t})$ or $v \in flatten(t^i \setminus \hat{t}) \cap flatten(t^j \setminus \hat{t})$.
	Then $rhist^i[v] = rhist^j[v]$ follows from Lemmas~\ref{lemma:t0-equal-rhist} and \ref{lemma:t-hat-equal-rhist}.
\end{IEEEproof}

Induction step 2 of Lemma~\ref{lemma:all-equal-rhist}: The fixed SSCC trace~$\hat{t' }$ has length $|\hat{t}|$ and the corresponding SCC trace~$t$ ends with a sequence of regular SCCs.
\begin{lemma}
	\label{lemma:tplus-equal-rhist}
	At replica~$r_i$ and $r_j$ for two traces~$t^i$ and $t^j$ with identical SSCCs, that is $\hat{t^i}=\hat{t^j}$: $\forall v\in flatten(t^i) \cap flatten(t^j): rhist^i[v] = rhist^j[v]$, where $s$ is an SCC.
\end{lemma}
\begin{IEEEproof}
	W.l.o.g. we assume $|t^i| = |t^j|$. A short trace can be padded with empty "SCCs" which correspond to a no-op.
	In addition, we assume that there is a position~$x$ such that $t^i[x] = \hat{t}[|\hat{t}-1|] = t ^j[x]$, that is the last SSCC is at the same index in both traces.
	By assumption, all SCCs after the SSCC are executed regularly.

	Base case: $|t| = x+1$: Follows from Lemma~\ref{lemma:t-hat-equal-slot-rhist}.
	
	Induction step: $|t'| = |t|+1$:
	We define $s^i := t'^i[|t'^i|-1]$ to be the last element in $t'^i$. We only discuss $s^i$, the same arguments apply for an $s^j$ with swapped $i$ and $j$.
	
	\textbf{Case 1}: $s^i \in set(t'^i) \cap set(t'^j)$:  The same as Case 1 in the proof of Lemma~\ref{lemma:t0-equal-weak-rhist}, except that the induction assumption is strengthened with Lemma~\ref{lemma:t-hat-equal-slot-rhist}.
	
	\textbf{Case 2}: $s^i \notin set(t'^i) \cap set(t'^j) \Leftrightarrow s^i \notin t'^j$: This never applies as regular SCCs and SSCCs are disjoint, according to Lemma~\ref{lemma:t0-equal-weak-rhist} (Case 2) and \ref{lemma:t-hat-equal-rhist}.
\end{IEEEproof}

We now finish the proof of Lemma~\ref{lemma:all-equal-rhist}.

\noindent\textbf{Lemma~\ref{lemma:all-equal-rhist} (repetition).} \emph{Assume replica~$r_i$ and $r_j$ have two traces~$t^i$ and $t^j$:$ \forall v\in flatten(t^i) \cap flatten(t^j)$. Then $rhist^i[v] = rhist^j[v]$, where $v$ is a slot in an SCC.}
\begin{IEEEproof}
	We assume w.l.o.g. that $|\hat{t^i}|=|\hat{t^j}|$.
	A short SSCC trace can be padded with empty no-op SSCCs.
	
	Base case: $|\hat{t^i}|=|\hat{t^j}|=0$: See Lemma~\ref{lemma:t0-equal-rhist}.
	
	Induction step: $|\hat{t^i}'| = |\hat{t^i}|+1$:
	We only show this for $\hat{s^i} := \hat{t^i}'[|\hat{t^i}'|-1]$, a symmetrical argument applies for $s^j$.
	
	\textbf{Case 1:} $\hat{s^i} \in set(\hat{t^i}') \cap set(\hat{t^j}') $:
	All SCCs on which $\hat{s^i}$ depends have already been executed, thus $rdeps_{exp}^i(\hat{s^i})=rdeps_{exp}^j(\hat{s^i})$ as shown in the proof of Lemma~\ref{lemma:t-hat-equal-rhist}.
	
	\textbf{Case 2:} $\hat{s^i} \notin set(\hat{t^i}') \cap set(\hat{t^j}') $:
	We show that if $\hat{s^i} \neq \hat{s^j}$, for a $\hat{s^j} \in \hat{t^j}$ then $\hat{s^i} \cap \hat{s^j} = \emptyset$.
	For that we show that an SSCC can be identified by a single slot~$v$.
	An SSCC depends on $\mathcal{D} := \bigcup_{v\in \hat{s^i}} exp(v.i)$.
	$v\in\mathcal{D}$ must be part of the SSCC as otherwise they would have been executed before.
	Thus, $exp(...)$ is defined by the SSCC.
	And therefore $\forall v \in \hat{s^i}: rdeps_{exp}^i(v)=rdeps_{exp}^j(v)$.
	
	The cases are exhaustive.
\end{IEEEproof}

\noindent\textbf{Theorem~\ref{thm:sscc-trace-completeness} (repetition)} (Execution Consistency)\textbf{.} \emph{All replicas execute all pairs of committed, conflicting requests in the same order.}

Now we proof the theorem:
\begin{IEEEproof}
	The agreement guarantees that for two conflicting requests $a$ and $b$ in slots $v_1$ and $v_2$, at least one will depend on the other.
	W.l.o.g. assume that $v_2 \in deps(v_1)$ and that $v_1$ and $v_2$ were already executed.
	
	\textbf{Case 1:} $v_1$ and $v_2$ are part of the same regular SCC or SSCC:
	An SCC is sorted before executing, thus ensuring a stable order.
	
	\textbf{Case 2:} $v_2 \in rhist[v_1]$:
	Then $v_2$ was executed before $v_1$. Assume this is not the case:
	This is only possible if $v_1$ and $v_2$ are part of a single SCC, which contradicts the assumption.
	
	\textbf{Case 3:} $v_2 \notin rhist[v_1]$:
	$v_1$ must be part of an SSCC, as only $rdeps_{exp}$ can exclude dependencies from $deps(v_1)$.
	
	We first show that the execution behaves as if the following additional dependencies for each slot exist: $\tilde{deps}(v) \hat{=} deps(v) \cup \{x\,|\,x.i=v.i \wedge x.seq \leq v.seq - k\}$.
	We adapt $\tilde{rdeps}$ and $\tilde{rdeps}_{exp}$ to use $\tilde{deps}(v)$.
	When $v$ is executed, $v\in exp_k$ thus $x.seq \leq v.seq - k < exp(v.i).seq$ and thus the additional dependencies only point to already executed slots.
	Therefore, they do not affect the execution.
	
	When the SSCC was executed $v_1 \in exp_k, v_2 \notin exp_k$.
	Then we get: $v_r = exp(v_2.i) \in deps(v_1)$ due to Corollary~\ref{corollary:compact-deps}, $v_r \in \tilde{deps}(v_1)$, $v_r \in \tilde{deps}(v_2)$ and $(v_1 \leadsto v_r \wedge v_r \leadsto v_1) \vee v_1 = v_r$ as $v_1$ and $v_r$ are part of the SSCC.
	Therefore, $v_2$ can only execute after the SSCC as $v_r \in SSCC$ and $v_2 \rightarrow v_r$.
	
	The cases are exhaustive.
	
	In contrast to the SCC trace, the execution pseudocode starts from individual slots and tests whether a slot and its dependency graph are executable.
	Only then the SCCs are calculated and executed.
	When the tested slots are part of the SCC to execute next, then it is trivial to see that both representations are equivalent.
	Now suppose slot~$v_b$ of SCC~$B$ which depends on SCC~$A$ is tested first.
	If both SCCs are regular SCCs, then SCC~$A$ will be executed before $B$.
	As $rhist[v_a] := rdeps(A)$ for $v_a \in A$ it makes no difference whether $rdeps(B)$ is calculated before or after executing SCC~$A$.
	
	If only $A$ is an SSCC, then $A$ is executed first and afterwards the execution is restarted, which includes a recalculation of $rdeps(v_b)$.
	If only $B$ is an SSCC, then we arrive at a contradiction, as $A$ must already have been executed.
	If both are SSCCs, then one of both is executed and afterwards the execution is restarted.
	In all these cases the behavior is equivalent to that assumed when working with SCC traces.
	This generalizes to dependency graphs which contain more than two SCCs.
\end{IEEEproof}

\begin{remark}
	It is sufficient for the unblock execution case to only check slots in $exp(*)$, that is the root nodes.
	As shown in Lemma~\ref{lemma:scc-sscc-uniqueness} at least one slot in every SSCC is $\in exp(*)$.
\end{remark}
\begin{remark}
$rhist$ can be ignored for an implementation, as by construction it only contains executed slots.
An already executed slot cannot have dependencies on not yet executed slots.
Therefore, slots in $rdeps(v)$ and $rdeps_{exp}(v)$ can be split into two sets $\mathcal{A} \subseteq executed$ and $\mathcal{B} \cap executed = \emptyset$ with executed and not executed slots, respectively.
Only slots in $\mathcal{B}$ can depend on slots in $\mathcal{A}$.
A similar structure applies for the SCCs in $rdeps(v)$ or $rdeps_{exp}(v)$.
As these SCCs are skipped if they were executed before, it is equivalent to remove executed slots from $rdeps$ or $rdeps_{exp}$ as well.
The simplest way to achieve that is to drop $rhist$ completely.
\end{remark}
\begin{remark}
	\label{remark:single-graph}
	An implementation can handle $rdeps$ and $rdeps_{exp}$ using a single graph and immediately remove executed slots.
	This is easy to see for $rdeps$ alone, the combination with $rdeps_{exp}$ requires small modifications:
	Only slots $\in exp_k$ should be processed, all other slots can be regarded as not yet committed.
	Then $rdeps(v) \not\subseteq committed \Leftrightarrow rdeps(v) \not\subseteq exp_k$.
	$rdeps_{exp}(v)$ can be emulated by ignoring dependencies on slots $\notin exp_k$ on the fly.
\end{remark}

\subsubsection{Linearizability}
\begin{theorem}[Linearizability]
	If two interfering reqeusts are proposed one after another, such that the first request is executed at some correct replica before the second request is proposed, then all replicas will execute the requests in that order.
\end{theorem}
\begin{IEEEproof}
	Once a request~$a$ was executed then all later conflicting requests~$b$ will depend on $a$ and are thus ordered after $a$.
	To prevent the duplicate execution of client requests, the requests of a client always conflict with each other, which guarantees a total order for the requests of each client.
\end{IEEEproof}

\subsubsection{Agreement Liveness}
Similar to the Liveness property in EPaxos~\cite{moraru13there}, we show:
\begin{theorem}[Agreement Liveness]
	\label{thm:agreement-liveness}
	In synchronous phases a client request will commit eventually.
\end{theorem}
We first show that dependencies proposed by correct replicas will be accepted eventually, then show that a slot will commit and finish by showing that this also holds for a client request.

\begin{definition}
We say that \texttt{wait}~(Line~\ref{code:agr-wait}) accepts a slot as dependency, if the function does not block permanently, that is it returns eventually.
\end{definition}

\begin{lemma}
	\label{lemma:eventual-correct-msgpropose}
	If a correct replica~$r_i$ has accepted a \msgpropose from replica~$r_j$, then all other correct replicas will accept it as a dependency eventually.
\end{lemma}
\begin{IEEEproof}
	We show this by induction:
	For the base case assume that the \msgpropose contains no dependencies.
	The propose timeout stays active at replica~$r_i$ until it has accepted $2f$~\msgverify{}s for the \msgpropose~(L.~\ref{code:agr-stop-propose-timeout}).
	\begin{itemize}
		\item \textbf{Case 1:} Coordinator~$r_j$ is correct.\\
		All replicas will receive the \msgpropose and thus \texttt{wait} accepts the slot as dependency.
		\item \textbf{Case 2:} Coordinator~$r_j$  is faulty.\\
		Assume that replica~$r_i$ has accepted $2f$~\msgverify{}s. Only $f-1$ faulty \msgverify{}s are possible. Thus, at least $f+1$ of $2f$ \msgverify{}s are from correct replicas.
		And therefore all replicas will receive $f+1$~\msgverify{}s causing wait to accept the dependency.
		Otherwise, the propose timeout will expire, causing a correct replica~$r_k$ to broadcast the request.
		This allows all other replicas to learn about the slot corresponding to the \msgpropose as the message was signed by the request coordinator.
		
		Alternatively, if replica~$r_i$ receives the \msgpropose, then it will broadcast the message if it fails to collect $2f$~\msgverify{}s within the propose timeout.
		\item \textbf{Case 3:} A view-change triggers at replica~$r_i$.\\
		The replica~$r_i$ broadcasts the \msgpropose to all replicas if the propose timeout was still active~(Line.~\ref{code:agr-propose-on-vc}).
		\item The cases are exhaustive.
	\end{itemize}
	
	For the induction step we look at a later \msgpropose for which the correct replica~$r_i$ must also have accepted all dependencies.
	Thus, if it is necessary to broadcast the \msgpropose it will be accepted eventually as all dependencies will be accepted due to the induction assumption. 
\end{IEEEproof}
\begin{remark}
	\label{remark:liveness-vc-propose-broadcast}
	Note that the view-change special case  to broadcast the \msgpropose~(Line.~\ref{code:agr-propose-on-vc}) only exists for completeness, but is not strictly necessary during synchronous phases.
	For a view change at least one correct replica~$r_k$ must have sent a \msgvc.
	This in turn requires that the replica~$r_k$ has either received the \msgpropose in which case it will broadcast the \msgpropose itself if necessary.
	Or the replica~$r_k$ has received $f+1$~valid \msgverify{}s in which case at least one of these was sent by a correct replica, which must have received the \msgpropose and therefore also ensures its distribution.
	Together with the commit timeout~$9\Delta$ which is much larger than the propose timeout of $2\Delta$, the special case can only trigger if another correct replica has received and possibly distributed the \msgpropose before.
\end{remark}

\begin{lemma}
	\label{lemma:eventual-accept-deps}
	A dependency included in a request proposed by a correct replica~$r_i$ will be accepted by all correct replicas eventually.
\end{lemma}
\begin{IEEEproof}
	By construction, correct replicas only propose dependencies, for which they accepted the corresponding~\msgpropose.
	Then according to Lemma~\ref{lemma:eventual-correct-msgpropose} the corresponding messages will be accepted (by \texttt{wait}).
\end{IEEEproof}

\begin{lemma}
	\label{lemma:wait-eventual-commit}
	\texttt{wait} only accepts slots as dependencies which will commit eventually.
\end{lemma}
\begin{IEEEproof}
	The \texttt{wait} function waits for each dependency until one of the following cases holds~(L.~\ref{code:agr-wait-start}-\ref{code:agr-wait-end}):
	\begin{itemize}
		\item \textbf{Case 1:} $f+1$~\msgverify{}s received.\\ These include at least one \msgverify from a correct replica, which must have received a valid \msgpropose and which will broadcast it if necessary.
		\item \textbf{Case 2:} $f+1$~\msgvc{}s received.\\ At least one \msgvc is from a correct replica, which also ensures that a correct replica has received a valid \msgpropose, see Remark~\ref{remark:liveness-vc-propose-broadcast}.
		\item \textbf{Case 3:} \msgpropose accepted.\\ This enables a replica to broadcast the \msgpropose itself if necessary.
		\item The cases are exhaustive.
	\end{itemize}
	Together with Lemma~\ref{lemma:eventual-correct-msgpropose} and \ref{lemma:eventual-accept-deps} eventually the commit timeout is active at all correct replicas which forces the slot to commit.
\end{IEEEproof}

Assume for now that the used timeout values are large enough to ensure progress.

\begin{lemma}
	\label{lemma:eventual-coorect-coord}
	A slot (not request) of a correct coordinator will commit eventually.
\end{lemma}

\begin{IEEEproof}
\textbf{Case 1:} The fast-path quorum~$F$ only contains correct replicas.

Then one of the following can happen:
	\begin{itemize}
\item Case 1.1: The slot commits without view change.
\begin{IEEEproof}
	Correct replicas enforce that a coordinator does not leave gaps in its sequence number space~(L.~\ref{code:agr-wait-no-gaps}).
	During  a synchronous phase, the \texttt{wait} calls in lines~\ref{code:agr-wait-no-gaps} and \ref{code:agr-wait-verify} do not block permanently according to Lemma~\ref{lemma:eventual-accept-deps}.
	The coordinator and the fast-path quorum make up a total of $2f+1$ correct replicas which allows the slot to commit.
	
	In an asynchronous phase the coordinator will retransmit its \msgpropose until all correct replicas have received it.
	Then either the slot will commit or $\geq f+1$~replicas trigger a view change.

	The replicas start the commit timeout after receiving the \msgpropose or in the case of the request coordinator after sending the \msgpropose and thus either commit or request a view change.
	Once $f+1$~correct replicas have committed, then the remaining $f$~correct replicas can only trigger a view change with the help of faulty replicas.
	When the view-change does not start within timeout~$\Delta_{query-exec}$ after sending the own \msgvc, then a replica issues \msgqueryexec requests to all other replicas~(L.~\ref{code:agr-broadcast-query-exec}).
	These up to $f$~replicas then receive the result via \msgexec messages from at least $f+1$ correct replicas.
\end{IEEEproof}
\item Case 1.2: A view change is necessary for at least one replica.
\begin{IEEEproof}
	As soon as $f+1$~correct replicas have issued a \msgvc for view~$v+1$  then eventually all correct replicas will issue a~\msgvc(L.~\ref{code:agr-vc-inc-start}-\ref{code:agr-vc-inc-end}).
	In a synchronous phase eventually all correct replicas will enter the view change in the same view~$v+1$.
	
	The timeout for view~$v+2$ is only started after ensuring that at least $f+1$~correct replicas have reached view~$v+1$ and sent a \msgvc. This in turn ensures that all correct replicas will reach view~$v+1$ at the same time if the network in synchronous, see also Lemma~\ref{lemma:new-view-timeout}.
	Then all correct replicas will start their view change timeout, as enough \msgvc{}s exist to ensure that a \msgnewview can be created eventually.
	Then either at least $f+1$~correct replicas accept the \msgnewview or $f+1$~correct replicas switch to view~$v+2$.
	
	After a replica~$r_i$ accepts a \msgnewview, then a different replica~$r_j$ will either eventually also receive and accept the \msgnewview or switch to a higher view.
	As $2f+1$~\msgvc{}s are necessary to compute a \msgnewview, at least $f+1$ must be from correct replicas, thus eventually all replicas start a view change, will receive $2f+1$~\msgvc and start their view-change timeouts.
	Then either the replica accepts the \msgnewview or switches to a higher view.
	After accepting a \msgnewview a replica restarts the commit timeout which again ensures that the reconciliation path completes or another view-change is started.
	
	The view-change coordinator is rotated in each view such that eventually a correct coordinator is used, which allows the slot to commit.
\end{IEEEproof}
\item Case 1.3: \msgpropose and \msgverify (from correct replicas) contain dependencies not accepted by \texttt{wait}.
\begin{IEEEproof}
	Using Lemma~\ref{lemma:eventual-accept-deps} we immediately arrive at a contradiction.
\end{IEEEproof}
\item The cases are exhaustive.
\end{itemize}

\textbf{Case 2:} Fast-path quorum~$F$ contains faulty replicas.

We show that faulty replicas in the fast-path quorum~$F$ cannot prevent committing a slot (only its request) and cannot add dependencies to  non-existing slots.
	The faulty replicas can exhibit one of the following behaviors:
	\begin{itemize}
		\item Case 2.1: A faulty replica sends multiple \msgverify{}s.
		\begin{IEEEproof}
			The replica can prevent the fast or reconciliation path from completing when replicas collect diverging or no $\vec{dv}$.
			If the faulty replica prevents the slots from committing then the commit timeout enforces a view change.
			This will result in filling the slot with \texttt{null} after the view change.
		\end{IEEEproof}
		\item Case 2.2: A faulty replica does not send a \msgverify.
		\begin{IEEEproof}
			Same as the previous case.
		\end{IEEEproof}
		\item Case 2.3: A faulty replica proposes non-existing dependencies.
		\begin{IEEEproof}
			According to Lemma~\ref{lemma:wait-eventual-commit}, correct replicas, which have received the \msgpropose, will time out while waiting for the dependencies to commit.
			This will trigger a view change which will fill the slot with \texttt{null}.
			Thus, non-existing dependencies for a slot cannot commit.
		\end{IEEEproof}
		\item The cases are exhaustive.
	\end{itemize}

The cases are exhaustive.
\end{IEEEproof}

\begin{lemma}
	\label{lemma:eventual-correct-fast-quorum}
	The fast-path quorum~$F$ will eventually contain only correct replicas.
\end{lemma}
\begin{IEEEproof}
	After filling a slot with \texttt{null} during the view change, the fast-path quorum is rotated.
	This will eventually result in the fast-path quorum~$F$ to only contain correct replicas.
\end{IEEEproof}

\begin{remark}
	Note that a faulty replica cannot prevent slots of correct replicas from committing by proposing manipulated \msgpropose{}s.
	Assume this were the case.
	Then a correct replica has to accept a \msgpropose from the faulty coordinator.
	Then Lemma~\ref{lemma:eventual-correct-msgpropose} applies, which yields a contradiction.
	Thus, a faulty coordinator can only cause its \msgpropose to block permanently in \texttt{wait} which will also prevent all further slots of the faulty replica from committing~(L.~\ref{code:agr-wait-no-gaps}) until the faulty \msgpropose is finally accepted.
\end{remark}

We now show that the timeout values are sufficient to ensure progress.
\begin{lemma}
	\label{lemma:wait-three-delta}
	A \msgpropose or \msgverify of a correct replica~$r_i$ will be accepted after at most $3\Delta$ after sending.
\end{lemma}
\begin{IEEEproof}
	Replica~$r_i$ has received the \msgpropose of a dependency as otherwise it would not include the dependency.
	The propose broadcast timeout is $2\Delta$.
	Thus, after $2\Delta$ replica~$r_i$ has either received $2f$~\msgverify{}s and thus after an additional $\Delta$ all replicas have received $f+1$~\msgverify{}s after which \texttt{wait} accepts the dependency.
	Or replica~$r_i$ broadcasts the \msgpropose which will reach all replicas within $\Delta$.
\end{IEEEproof}

\begin{lemma}
	\label{lemma:new-view-timeout}
	The calculation of a \msgnewview can complete within at most $3\Delta$ in synchronous phases.
\end{lemma}
\begin{IEEEproof}
	Once a correct replicas has received $2f+1$ \msgvc{}s then within $2\Delta$ every correct replica will receive $2f+1$ \msgvc.
	This allows the view-change coordinator to calculate the \msgnewview which after $\Delta$ arrives at all replicas.
	That is, in total a timeout of $3\Delta$ is sufficient.
\end{IEEEproof}

\begin{lemma}
	A commit timeout of at least $8\Delta$ allows correct coordinators to commit in synchronous phases.
\end{lemma}
\begin{IEEEproof}
	It can take $3\Delta$ each until a \msgpropose and \msgverify are accepted.
	The fast path takes another $\Delta$ until \msgfpcommit  reaches all replicas.
	On the reconciliation path \msgprepare and \msgcommit require up to $2\Delta$.
	This yields a total timeout of $8\Delta$. 
	As the timeout cannot start before the \msgpropose was sent, this is sufficient in all cases.
	
	After a view change $\Delta_{vc-commit} = 3\Delta$ is sufficient as the reconciliation path only requires up to $2\Delta$ and the receipt time of a correct \msgnewview can only vary by $\Delta$ between replicas. 
\end{IEEEproof}

\noindent\textbf{Theorem~\ref{thm:agreement-liveness} (repetition)} (Agreement Liveness)\textbf{.} \emph{In synchronous phases a client request will commit eventually.}

Now, we show Theorem~\ref{thm:agreement-liveness}:
\begin{IEEEproof}
	For slots in which the request was replaced by \texttt{null} the request coordinator will propose the request again~(L.~\ref{code:agr-repropose-after-null}).
	Together with Lemmas~\ref{lemma:eventual-coorect-coord} and \ref{lemma:eventual-correct-fast-quorum} this ensures that a slot / slots and also eventually a request will commit.
	The client also broadcasts its request after a timeout to all replicas.
	This ensures that a correct coordinator receives the request and commits it.
\end{IEEEproof}

\begin{lemma}
	The compact dependency representation does not break Liveness.
\end{lemma}
\begin{IEEEproof}
	The additional dependencies to replica $r_j$ have sequence numbers $s_d$ which are lower than the maximum sequence number~$max_{s_j}$ to which an explicit dependency exists.
	That is $s_d < max_{s_j} = \max_{r_j}\{d \in D_i \,|\,d.i = r_j \}$.
	A correct replica accepts a \msgpropose for $max_{s_j}$ only if it has seen all earlier sequence numbers, that is \texttt{wait} must already have accepted these~(L.~\ref{code:agr-wait-no-gaps}).
	Thus the guarantees provided by \texttt{wait} also include the earlier additional sequence numbers~$s_d$.

	The dependency representation does not affect execution consistency as it can only add, but not remove, dependencies.
\end{IEEEproof}

\subsubsection{Execution Liveness}
\begin{theorem}[Execution Liveness]
	In synchronous phases a client will eventually receive a result.
\end{theorem}

\begin{lemma}
	\label{lemma:eventual-dependencies-committed}
	Any slot included as dependency of a committed slot will commit eventually.
\end{lemma}
\begin{IEEEproof}
	The \texttt{wait} calls in Lines~\ref{code:agr-wait-no-gaps} and \ref{code:agr-wait-verify} together with Lemma~\ref{lemma:wait-eventual-commit} ensure that all dependencies of any committed slot will commit eventually.
\end{IEEEproof}

\begin{lemma}
	\label{lemma:eventual-execute}
	A committed request will be executed eventually.
\end{lemma}
\begin{IEEEproof}
	Lemma~\ref{lemma:eventual-dependencies-committed} shows that all slots on which a committed slot depends will commit eventually.
	In order to avoid the execution livelock problem discussed in EPaxos~\cite{moraru13there}, we now show that there is a finite upper bound for the number of slots which have to commit before a slot can be executed.
	
	A slot~$s$ can be executed via the normal case if all slots in $rdeps(s) \subseteq exp_k$.
	As $exp_k$ by construction only includes up to $k$ not executed slots per replica, the number of dependee slots is bounded.
	
	In addition, the unblock execution case executes slots in $rdeps_{exp}(s)$ which by construction always  is $\subseteq exp_k$ and thus all slots executed via the unblock case also only have to wait for a bounded number of dependencies.
	
	A slot~$s$ can only depend on a bounded number of slots (as all dependencies must have been proposed before).
	Thus, if any dependency~$d$ among these dependencies is not yet executed and therefore can prevent execution of $s$, then it serves as an upper bound for $exp(d.i).seq \leq d.seq$.
	Other dependee slots can add further upper bounds on $exp(*)$ which restrict the size of the dependency set even further.
	As the lowest upper bound per replica is relevant, a dependency chain can only include additional requests by depending on another replica which is not yet part of $rdeps(s)$ or $rdeps_{exp}(s)$.
	As the number of replicas is fixed, this can only add dependencies to a bounded number of slots.
\end{IEEEproof}
The theorem follows by combining Theorem~\ref{thm:agreement-liveness} and Lemma~\ref{lemma:eventual-execute}.

\subsubsection{Checkpoint Correctness}
We now extend the proof and pseudocode to also include the checkpointing mechanism of \system.

We only show the modified parts of the agreement and execution pseudocode below.
Grey lines are unchanged.
\vspace{3mm}\hrule\vspace{-1mm}
\def\lstbsty{\linespread{1.1}\footnotesize\lsttt}
\def\lstbstygrey{\lstbsty\color{gray}}
\def\nolstskip{}
\begin{lstlisting}[name=pseudocode,belowskip=0pt]
%\pseudocode{\textbf{Variables at each replica}}%:
$\Delta_{vc} := 5 \Delta$%\vspace{2mm}\hrule\begin{center}\textbf{Fast Path}\end{center}\hrule\vspace{2mm}%
%\pseudocode{Propose checkpoint request} \msgcpreq\pseudocode{if}% $s_j.seq~\textrm{mod}~cp\_interval = 0$
\end{lstlisting}
\begin{lstlisting}[name=pseudocode,basicstyle=\lstbstygrey,aboveskip=0pt,belowskip=0pt]
%\pseudocode{\textbf{Follower} }%$f_i$%\pseudocode{ \textbf{receives} }%$dp := \langle \langle \msgpropose{}, s_j, co, h(r), D, F\rangle, r \rangle$:
	%\lstbtt{pre:}% $step[s_j] =$ init
\end{lstlisting}
\begin{lstlisting}[name=pseudocode,aboveskip=0pt,belowskip=0pt]
	assert $(s_j.seq~\textrm{mod}~cp\_interval = 0) \oplus (r = \msgcpreq)$ /* Each replica must propose a checkpoint request exactly every cp_interval slots%\,%*/
\end{lstlisting}
\begin{lstlisting}[name=pseudocode,basicstyle=\lstbstygrey,aboveskip=0pt,belowskip=0pt]
	[...]
%\textbf{conflicts($r$):}%
\end{lstlisting}
\begin{lstlisting}[name=pseudocode,aboveskip=0pt]
	/* A checkpoint request $\msgcpreq$ conflicts with all other requests%\,%*/
	%\pseudocode{Return }%$\{s_i |\forall s_i, pr[s_i] \neq \emptyset: conflict(pr[s_i], r) \}$ $\cup$ %\pseudocode{barrier of latest stable checkpoint}%
\end{lstlisting}
\hrule\begin{center}\footnotesize\textbf{View Change}\end{center}\hrule
\begin{lstlisting}[name=pseudocode,basicstyle=\lstbstygrey,belowskip=0pt]
%\pseudocode{\textbf{Move to new view} }%$v_{s_j}$%\pseudocode{ \textbf{for slot} }%$s_j$%\pseudocode{ \textbf{at replica} }%$r_i$:
	[...]
	%\pseudocode{Else If }%$step[s_j] \in$ {rp-prepared, rp-committed}:
		[...]
\end{lstlisting}
\begin{lstlisting}[name=pseudocode,aboveskip=0pt,belowskip=0pt]
	%\pseudocode{Else If }%$s_j.seq~\textrm{mod}~cp\_interval = 0$:
		$D_{f_i} := D_{f_i} $%\pseudocode{used by $r_i$ for own \msgpropose{} / \msgverify or as fallback }%$ conflicts(\msgcpreq) \setminus s_j$
		$dv := \langle \msgverify{}, s_j, f_i, h(\msgcpreq), D_{f_i}\rangle _{\sigma_{f_i}}$
		$cert[s_j] := \langle \textsc{CRC-part}, \msgcpreq, dv , -1 \rangle$ /* for view -1%\,%/
\end{lstlisting}
\begin{lstlisting}[name=pseudocode,basicstyle=\lstbstygrey,aboveskip=0pt,belowskip=0pt]
	$step[s_j] =$ view-change
	[...]
%\pseudocode{\textbf{VC-Coordinator} }%$co$%\pseudocode{ \textbf{for view} }%$v_{s_j}$%\pseudocode{ \textbf{receives valid} }%$VCS:=\{\langle \msgvc, v_{s_j}, s_j, *, *\rangle\}$%\pseudocode{ \textbf{from} }%$2f+1$%\pseudocode{ \textbf{replicas}}%:
\end{lstlisting}
\begin{lstlisting}[name=pseudocode,aboveskip=0pt,belowskip=0pt]
	/* A $VC \in VCS$ containing a %\textsc{CRC-part}% is only considered valid after $wait(VC.dv.D_{f_i})$ has returned */%\label{code:agr-cp-crc-wait}%
\end{lstlisting}
\begin{lstlisting}[name=pseudocode,basicstyle=\lstbstygrey,aboveskip=0pt,belowskip=0pt]
	[...]
	%\pseudocode{Pick $dp$, $\vec{dv}$ from [...]}%
\end{lstlisting}
\begin{lstlisting}[name=pseudocode,aboveskip=0pt,belowskip=0pt]
	%\pseudocode{If}% $s_j.seq~\textrm{mod}~cp\_interval = 0 \wedge dp = null$:%\label{code:agr-cp-crc-start}%
		$dp := \msgcpreq$
		$\vec{dv} := \{ VC.dv\,|\,VC \in VCS  \}$ /* Each VC must contain a $\msgverify$%\,%*/%\label{code:agr-cp-crc-end}%
\end{lstlisting}
\begin{lstlisting}[name=pseudocode,basicstyle=\lstbstygrey,aboveskip=0pt]
	%\pseudocode{Broadcast }%$\langle \msgnewview{}, v_{s_j}, s_j, co, dp, \vec{dv}, VCS\rangle _{\sigma_{co}}$%\smallskip%
\end{lstlisting}
\hrule\vspace{5mm}
The execution pseudocode is adapted as follows:
\vspace{1mm}\hrule\vspace{-1mm}
\begin{lstlisting}[name=pseudocode,basicstyle=\lstbstygrey,belowskip=0pt]
execute($\vec{v}$, $d$):
\end{lstlisting}
\begin{lstlisting}[name=pseudocode,aboveskip=0pt,belowskip=0pt]
	$barrier := \emptyset$
	%\pseudocode{If }%$\msgcpreq \in \vec{v}$:
		$barrier := (\{x\,|\,\forall r_i: x < exp(r_i)\}$%\label{code:exec-cp-barrier}%$\bigcup_{v \in \vec{v}, v.req = \msgcpreq} deps(v) \cup v) \cap exp_k$
\end{lstlisting}
\begin{lstlisting}[name=pseudocode,basicstyle=\lstbstygrey,aboveskip=0pt,belowskip=0pt]
	%\pseudocode{For}% $c \in sort(\vec{v})$:
\end{lstlisting}
\begin{lstlisting}[name=pseudocode,aboveskip=0pt,belowskip=0pt]
		%\pseudocode{If}% $c \neq \msgcpreq \wedge (barrier = \emptyset \vee c \in barrier)$:
\end{lstlisting}
\begin{lstlisting}[name=pseudocode,basicstyle=\lstbstygrey,aboveskip=0pt,belowskip=0pt]
			%\pseudocode{Execute request}% $c$
			$rhist[v] := d$
\end{lstlisting}
\begin{lstlisting}[name=pseudocode,aboveskip=0pt]
	%\pseudocode{If}% $barrier \neq \emptyset$:
		%\pseudocode{Create execution checkpoint with}% $barrier$
		%\pseudocode{Restart request execution}%
\end{lstlisting}
\hrule\vspace{1mm}

As described in Section~\ref{sec:checkpointing} a replica broadcast a \msgcp message after creating a checkpoint.
Once a valid checkpoint is backed by at least $2f+1$~replicas, it becomes \emph{stable}.
This guarantees that the checkpoint is correct.
To apply a checkpoint, a replica requests the set of $2f+1$ checkpoint messages along with the checkpoint content and applies the checkpoint after verifying the correctness of all messages.

The Validity and Consistency properties are not affected by applying a checkpoint as this does not affect agreement slots except by garbage collecting old ones.
The Execution Consistency is also maintained as the execution state of a correct replica is applied.
As soon as a correct replica has a stable checkpoint, all other replicas will eventually be able to learn about the checkpoint.
This in turn allows all correct replicas to update their state if necessary.

The following Lemma adapts the proof of Theorem~\ref{thm:consistency} to handle checkpoint requests.
\begin{lemma}
	For a checkpoint slot, if a checkpoint request certificate~(CRC) is selected during a view-change then the slot did not commit previously.
\end{lemma}
\begin{IEEEproof}
	As shown in the proof of Theorem~\ref{thm:consistency} the new-view calculation always includes an FPC or RPC if either of both committed.
	Thus the CRC cannot be selected.
\end{IEEEproof}

\begin{lemma}
	All correct replicas create identical checkpoints for each checkpoint request.
\end{lemma}
\begin{IEEEproof}
	A checkpoint request conflicts with all other requests.
	This ensures that each request is either executed before or after the checkpoint request at all replicas due to the Execution Consistency property.
	In addition, this guarantees that all replicas execute a checkpoint request as part of the same SCC.
	Thus, all correct replicas execute the same part of the SCC before creating a checkpoint.
	As all replicas execute the same set of requests before a checkpoint, $exp_k$ is identical across replicas, and therefore all replicas bound the checkpoint barrier to the same slots~(Line~\ref{code:exec-cp-barrier}).

	We now show that the checkpoint barrier is tight.
	Assume that a slot~$x$ before the checkpoint barrier was not executed.
	$exp(*)$ which is added as lower bound to the checkpoint barrier cannot add unexecuted slots.
	For a slot~$x$ to be covered by the checkpoint barrier, the checkpoint request must include a dependency on $x$ or a slot $x' > x$.
	Then by Corollary~\ref{corollary:compact-deps} the checkpoint request depends on $x$ which therefore must be executed first.
	
	Assume that a slot~$x$ not covered by the checkpoint barrier was already executed.
	That slot must have been executed as part of a regular SCC or an SSCC.
	
	Assume that slot~$x$ was executed as part of an SSCC.
	The SSCC consists of at least 2 slots and therefore includes a dependency on the slot~$x$.
	Therefore, the SSCC also depends all slots between $exp(x.i)$ and the slot~$x$.
	Thus, after execution of the SSCC $exp(x.i) > x$.
	This yields a contradiction as the lower bound of the checkpoint barrier covers $\{x'\,|\,x' < exp(*)\}$.
	
	Assume slot~$x$ was executed as part of a regular SCC.
	$x$ must either depend on the checkpoint request or vice versa.
	When the  checkpoint request depends on $x$, it also depends on all slot between $exp(x.i)$ and $x$.
	Therefore $exp(x.i) > x$ when the checkpoint is executed, which yields a contradiction.
	Now, assume $x$ depends on the checkpoint request.
	Then $x$ must be executed after the checkpoint or as part of an SSCC, which both yields a contradiction.
	
	Thus all replicas create a checkpoint after executing the exact same set of requests.
	Applying the checkpoint yields the same state as a replica has after executing all requests up to the checkpoint.
\end{IEEEproof}
We now show that applying a checkpoint or garbage collecting slots after a checkpoint is stable does not affect Execution Consistency.
\begin{IEEEproof}
	Once a checkpoint is stable, all later requests will include dependencies on all slots included in the checkpoint.
	Compared to an execution without checkpointing this can only introduce additional dependencies.
	However, as all slots included in the checkpoint are already executed, these have no influence on the request execution.
\end{IEEEproof}

The following Lemma adapts the proof of Theorem~\ref{thm:agreement-liveness} to handle checkpoint requests.
\begin{lemma}
	For a checkpoint slot, if the slot did not commit previously then at least a checkpoint request certificate~(CRC) is selected during a view change.
\end{lemma}
\begin{IEEEproof}
	The new-view calculation requires $2f+1$~\msgvc{}s which are sufficient to generate a CRC~(Line \ref{code:agr-cp-crc-start}-\ref{code:agr-cp-crc-end}).
	As shown in the proof of Lemma~\ref{lemma:eventual-coorect-coord}, eventually all correct replicas will send a \msgvc.
	These messages and their dependencies will eventually be accepted by \texttt{wait} allowing the view change to complete~(Line~\ref{code:agr-cp-crc-wait}).
	Once a CRC has committed via the reconciliation path, then it is handled as any other request.
\end{IEEEproof}
We modify Lemma~\ref{lemma:new-view-timeout} as follows:
\begin{lemma}
	The calculation of a \msgnewview completes for a timeout of $5\Delta$ in synchronous phases.
\end{lemma}
\begin{IEEEproof}
	Once a correct replica has received $2f+1$ \msgvc{}s then within $2\Delta$ every correct replica will receive $2f+1$ \msgvc{}s.
	All \msgvc{}s are sent after $\Delta$ and are accepted at most $3\Delta$ later, similar to Lemma~\ref{lemma:wait-three-delta}.
	This allows the view-change coordinator to calculate the \msgnewview which after $\Delta$ arrives at all replicas.
	That is, in total a timeout of $5\Delta$ is sufficient.
\end{IEEEproof}

\subsubsection{Spatial Complexity}
The primary source of memory usage in \system are the agreement slots.
If each coordinator maintains $2 * cp\_interval$~slots, this yields a total of  $N * cp\_interval * 2$~slots.
For each slot, \system has to store the attached request and the messages necessary to create a fast-path or reconciliation-path certificate.
The dependency tracking in the agreement only has to track dependencies to slots explicitly contained in the compact dependency set, as that implicitly ensures that all earlier slots will also commit eventually.
Thus, only up to $N$~dependencies are tracked for a slot in the agreement.
That is the memory usage is roughly determined by $O(N * cp\_interval * (|r| + N*signature\_size))$.
This is similar to other protocols which forward certificates during the view-change.

As the execution only processes a window of $k$~agreement slots for each coordinator and only a single dependency graph is necessary as described in Remark~\ref{remark:single-graph}, the dependency graph in the execution only contains nodes for up to $N * k$~slots.
For each slot the request itself is still stored by the agreement, as otherwise it would have been garbage-collected from both the agreement and execution, and thus requires no additional memory.
The execution also has to unroll the compact dependency sets and add explicit edges between requests to the dependency graph.
Dependencies on already executed slots are not necessary for determining the execution order and thus are not stored.
Dependencies to slots beyond the window of currently executed requests, can be expanded once these slots enter the window.
This yields an upper bound of  $(N*k)^2$ for the number of edges.
The algorithm to calculate strongly-connected components uses a stack which in the worst case can contain all slots in the graph, that is up to $N * k$~elements.
In our evaluation, we have used $N=4$ and $k=20$, which limits the number of edges to a few thousands.
For these parameters, the memory usage for the execution is negligible compared to that for the agreement slots.

\end{document}